%% file: paper.tex
\documentclass[iop]{emulateapj}

\usepackage[normalem]{ulem}
\usepackage{color}
\usepackage{appendix}
\usepackage{textcomp}
\usepackage{gensymb}
\usepackage{breakcites}

\newcommand{\kgsins}[1]{\textcolor{black}{#1}}
\newcommand{\kgsdel}[1]{{}}

\usepackage[export]{adjustbox}
\usepackage{graphics}
\usepackage{hyperref}

\hypersetup{
    pdfauthor=Stefan Laos,
    pdftitle=Work in Progress,
    colorlinks=true,
    citecolor=blue,
    linkcolor=black,
    urlcolor=cyan}

\graphicspath{{Figures/}}

\shorttitle{Assessing Spectroscopic-Binary Multiplicity Properties Using Robo-AO Imaging}
\shortauthors{Laos, Stassun, \& Mathieu}

\begin{document}


\title{Assessing Spectroscopic--Binary Multiplicity Properties Using Robo-AO Imaging}


\author{Stefan Laos\altaffilmark{1}, Keivan G.\ Stassun\altaffilmark{1, 2}, and Robert D.\ Mathieu\altaffilmark{3}}

%\affiliation{Physics and Astronomy Department, Vanderbilt University, Nashville, TN 37203}
\altaffiltext{1}{Department of Physics and Astronomy, Vanderbilt University, Nashville, TN 37235, USA}
\altaffiltext{2}{Department of Physics, Fisk University, Nashville, TN 37208, USA}
\altaffiltext{3}{Department of Astronomy, University of Wisconsin--Madison, Madison, WI 53706, USA}

% Abstract

\input{Section_Files/Abstract.tex}

% Introduction
\input{Section_Files/Introduction.tex}

\input{Section_Files/Binary_Sample.tex}

\input{Section_Files/Imaging.tex}

\input{Section_Files/Companion.tex}

\input{Section_Files/Results.tex}

\input{Section_Files/Discussion_summary.tex}

\input{Section_Files/Acknowledgements.tex}
\bibliographystyle{aasjournal}
\bibliography{RoboAO.bib}
\nocite{*}

\clearpage

\begin{table*}
\centering
\caption{Robo-AO Non-SB Observations} \label{tab:RobonotSBs} 
\begin{tabular}{cccccccc}
\hline \hline
Name & R.A. (deg.) & Dec. (deg.) & V (mag) & Strehl Ratio (\%) & Multiplicity Flag\tablenotemark{a} \\
\hline
2MASS J05130342+2423489 & 78.264217899 & 24.396841655 & 6.707 & 2.06 & T \\
2MASS J08140761+3145095\tablenotemark{b} & 123.531723248 & 31.752562931 & 12.44 & 6.52 &  \\
2MASS J08145689+3208572 & 123.736957886 & 32.149168567 & 8.3 & 4.51 & B \\
2MASS J09303285+2735099 & 142.636911616 & 27.585936424 & 10.722 & 3.56 & N \\
2MASS J10404394+2348227 & 160.18296347 & 23.806274661 & 10.382 & 2.3 & N \\
2MASS J11040722+4415409 & 166.029847747 & 44.261378235 & 9.685 & 3.66 & B \\
2MASS J11464736+3424349 & 176.69737612 & 34.409630054 & 10.378 & 2.14 & N \\
2MASS J13312997+3813115 & 202.874932285 & 38.219833319 & 10.839 & 2.86 & N \\
2MASS J13334365+3754543 & 203.431735917 & 37.915035417 & 10.219 & 1.96 & N \\
2MASS J13392768+2726135\tablenotemark{b} & 204.864597905 & 27.437240111 & 11.777 & 6.72 &  \\
2MASS J15055682+2216202 & 226.486763946 & 22.27226675 & 8.71 & 3.36 & B \\
2MASS J15120615+6826579 & 228.025462449 & 68.449601426 & 9.621 & 1.68 & N \\
2MASS J16385624+3652029 & 249.734407561 & 36.867416544 & 10.311 & 2.72 & N \\
2MASS J17205010+4223016 & 260.208784546 & 42.38376309 & 10.75 & 3.87 & B \\
2MASS J19303059+3751364 & 292.627494341 & 37.860165321 & 10.688 & 4.86 & B \\
2MASS J19344300+4651099 & 293.6792058 & 46.852733735 & 7.768 & 1.26 & N \\
\hline
\end{tabular}
\footnotetext[1]{Multiplicity flag: B = binary, T = triple, N = no companion detected.}
\footnotetext[2]{Faint systems that were not reduced properly (see Section 2.1) (no companion information could be deduced from RoboAO imaging.)}
\end{table*}

\appendix
\section{Mosaic of PSF-subtracted RoboAO Images}

To display the companions identified from our RoboAO imaging, we first show a mosaic of our pre-PSF-subtracted images for each SB system, followed by a mosaic of our PSF-subtracted images with visual aids pointing to each resolved companion.

\begin{figure}[!ht]
    \includegraphics[width=\linewidth]{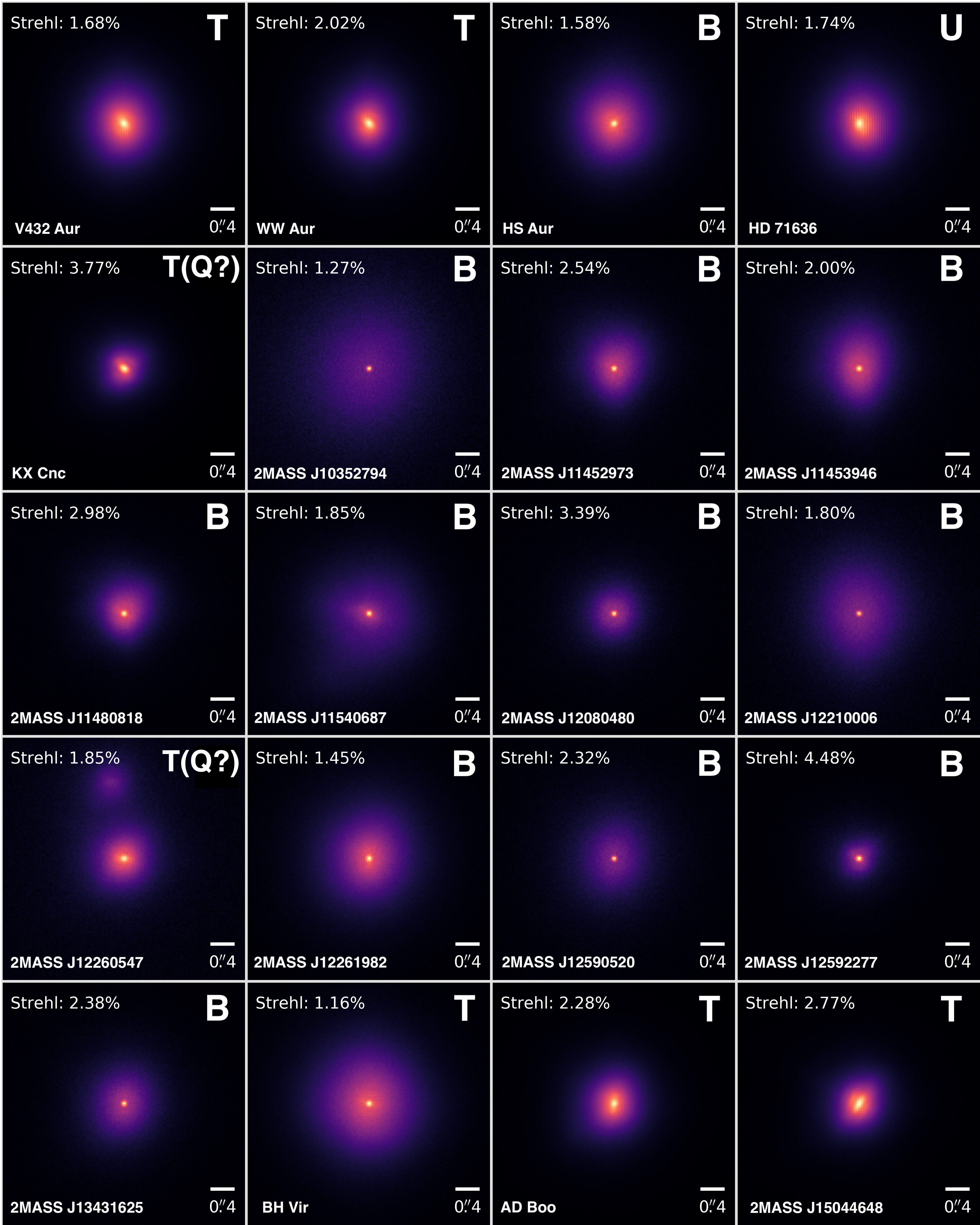}
    \caption{Pre-PSF-subtracted images for RoboAO SBs (continued in Figure~\ref{fig:PSFpresubmosaic2} and Figure~\ref{fig:PSFpresubmosaic3}). We note the designation in the top right corner of each image corresponds to that determined strictly by our RoboAO imaging and thus does not include the results of our wide companion identifications in Section 3. Our final identifications are that listed in Table 1.}
    \label{fig:PSFpresubmosaic1}
\end{figure}

\begin{figure}[!ht]
    \includegraphics[width=\linewidth]{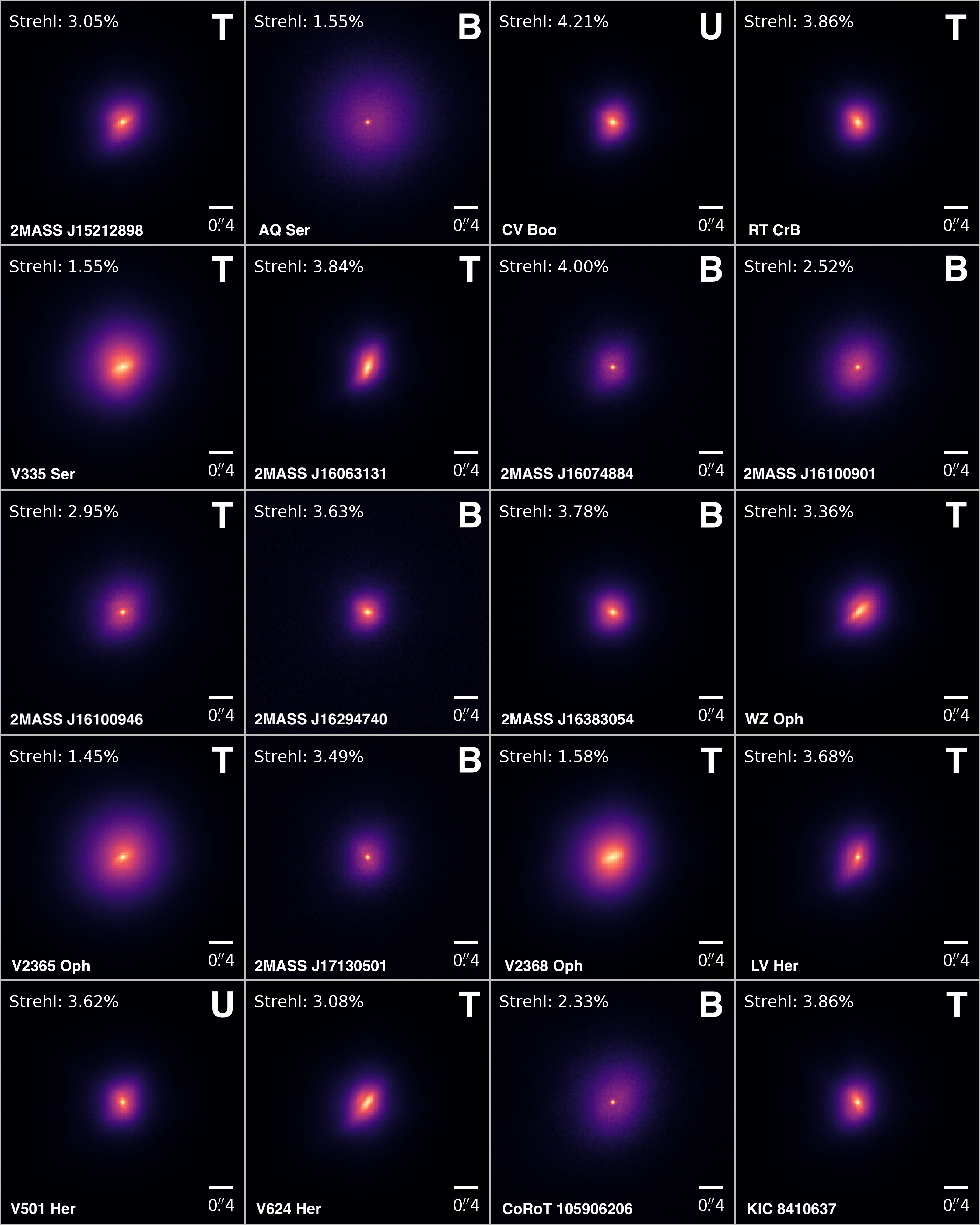}
    \caption{Pre-PSF-subtracted images for RoboAO SBs (continued from Figure~\ref{fig:PSFpresubmosaic1}). We note the designation in the top right corner of each image corresponds to that determined strictly by our RoboAO imaging and thus does not include the results of our wide companion identifications in Section 3. Our final identifications are that listed in Table 1.}
    \label{fig:PSFpresubmosaic2}
\end{figure}

\begin{figure}[!ht]
    \includegraphics[width=\linewidth]{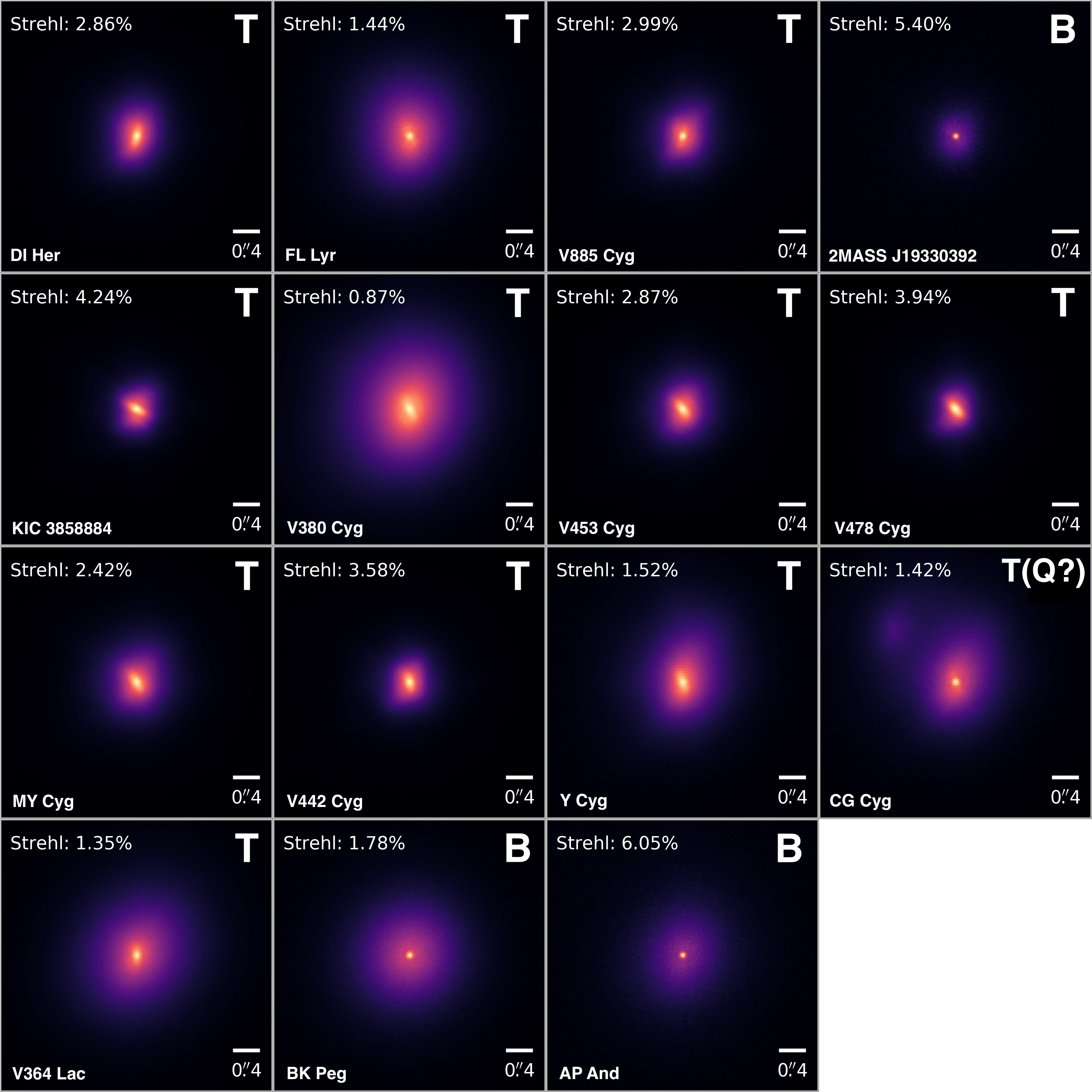}
    \caption{Pre-PSF-subtracted images for RoboAO SBs (continued from Figure~\ref{fig:PSFpresubmosaic2}). We note the designation in the top right corner of each image corresponds to that determined strictly by our RoboAO imaging and thus does not include the results of our wide companion identifications in Section 3. Our final identifications are that listed in Table 1.}
    \label{fig:PSFpresubmosaic3}
\end{figure}

\begin{figure}[!ht]
    \includegraphics[width=\linewidth]{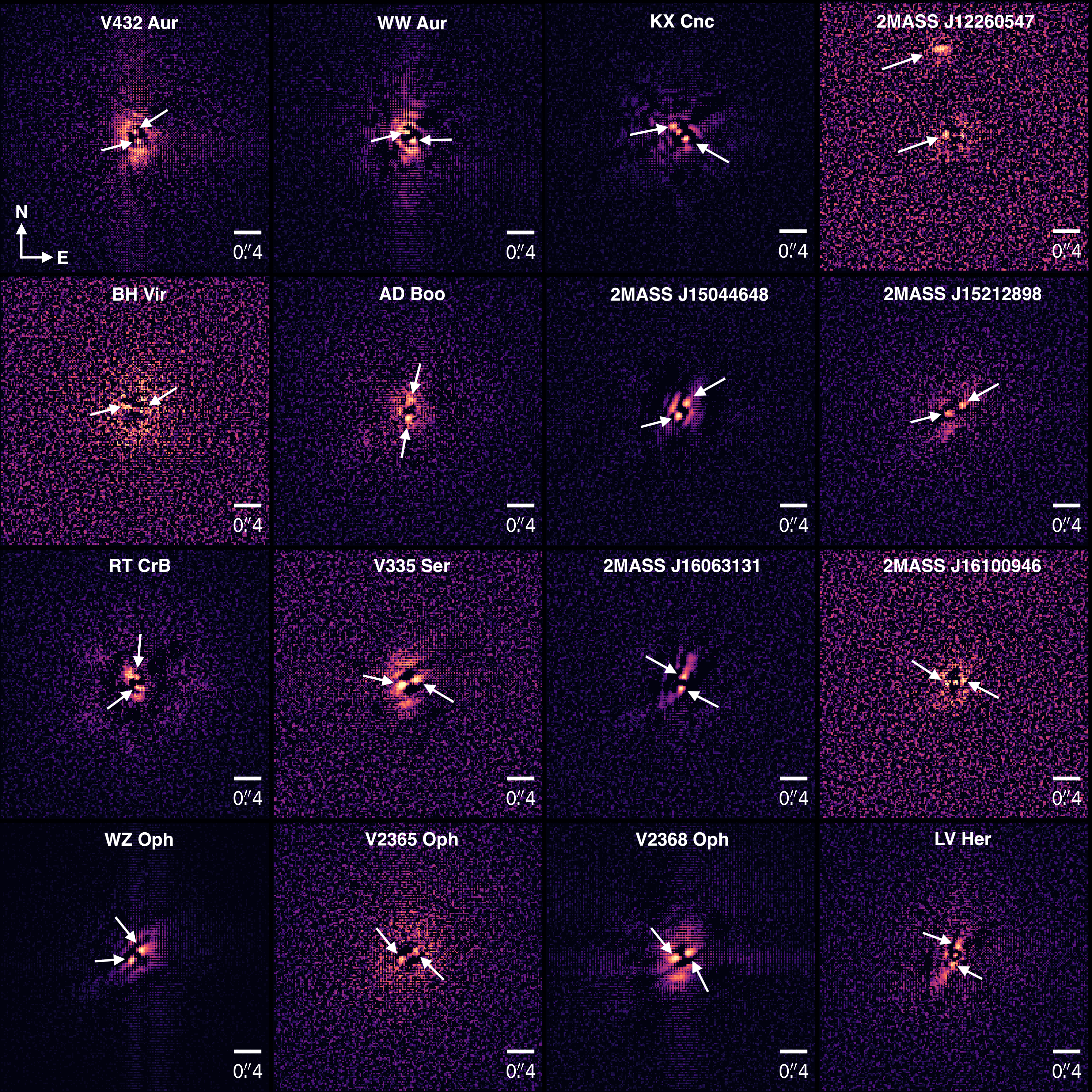}
    \caption{PSF-subtracted images for RoboAO identified multiples (continued in Figure~\ref{fig:PSFsubmosaic2}).} 
    \label{fig:PSFsubmosaic1}
\end{figure}

\begin{figure}[!ht]
    \includegraphics[width=\linewidth]{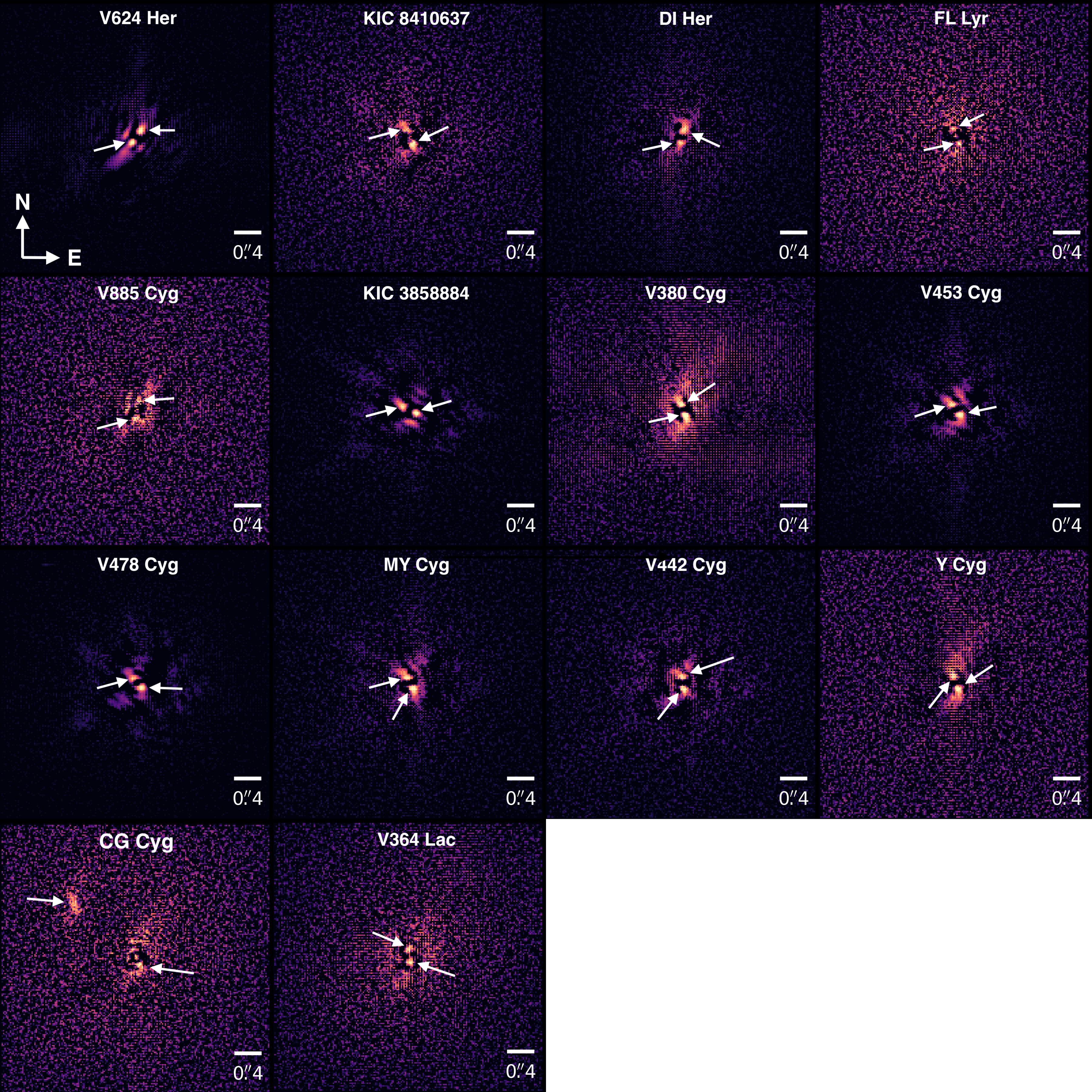}
    \caption{PSF-subtracted images for RoboAO identified multiples (continued from Figure~\ref{fig:PSFsubmosaic1}).}
    \label{fig:PSFsubmosaic2}
\end{figure}

\end{document}

%% file: Section_Files/Abstract.tex
\begin{abstract}
We present higher-order multiplicity results for 60 solar-type spectroscopic binaries based on 0.75$\mu$m imaging data taken by the Robotic Adaptive Optics system (Robo-AO) at the Kitt Peak 2.1m telescope. Our contrast curves show sensitivity of up to $\sim$5 mag at $\sim$1\arcsec\ separation. We find tertiary companions for 62\% of our binaries
overall, but find this fraction is a strong function of the inner binary orbital period; it ranges from $\sim$47\% for $P_{\rm bin}> 30^{d}$ to as high as $\sim$90\% for $P_{\rm bin} \lesssim 5^{d}$. 
We similarly find increasing tertiary companion frequency for shorter period binaries in a secondary sample of {\it Kepler\/} eclipsing binaries observed by Robo-AO. Using {\it Gaia\/} distances, we estimate an upper limit orbital period for each tertiary companion and
compare the tertiary-to-binary period ratios for systems in the field versus those in star-forming regions. Taken all together, these results provide further evidence for angular momentum transfer from three-body interactions, resulting in tight binaries with tertiaries that widen from pre-main-sequence to field ages.
\end{abstract}

%% file: Section_Files/Introduction.tex
\section{Introduction}

%{\bf [Stefan, take a fresh look at the Introduction from the standpoint of setting up (a) our goal of testing the reproducibility of the Tokovinin result, and (b) exploring the potential evolutionary pathways for multiple-star systems.]}

Although many of the underlying mechanisms and outcomes of the star-formation process remain under debate, the prevalence of stellar multiplicity 
%in nearby star forming regions 
is undisputed, with more than half of all stars having at least one companion. In the past decade, stellar multiplicity studies \citep[e.g.,][]{Raghavan2010,Sana2014,2014AJ....147...86T,Fuhrmann2017} have focused on volume-limited samples in distinct spectral-type ranges to achieve unbiased statistical inferences of multiple systems. Others \citep[e.g.,][]{Rucinski2007,Riddle2015,Hillenbrand2018} have used high-resolution, adaptive-optics imaging to identify and characterize new systems. 

In particular, high spatial resolution campaigns at longer wavelengths have the ability to maximize detections of faint distant companions. \citet{Tokovinin2006} surveyed 165 solar-type spectroscopic binaries (SBs) and found that virtually all ($\sim$96\%) short-period binary systems ($P_{\rm bin} <$ 3$^{d}$) had tertiary companions. 

\kgsins{This seminal finding has motivated a number of follow-on studies to explore the potential effects of, as well as the evolutionary pathways of, binaries in tertiary systems.}
The dynamics introduced by \kgsins{tertiaries could} have significant consequences for the evolution of young stellar objects. \kgsins{For example,} in the most recent review of benchmark pre-main-sequence (PMS) eclipsing binaries (EBs), \citet{Stassun2014} found that the properties of those in triple systems constituted the most highly discrepant cases when compared to PMS stellar evolution models, \kgsins{which} may be explained \kgsins{if} the tertiary inputs significant energy into one or both binary stars during periastron passages \kgsins{\citep[see, e.g.,][]{GomezMaqueoChew2019}}. 

\kgsins{More generally,} the dominant physical mechanism by which the tertiary influence\kgsins{s} the creation of tightly bound binaries \kgsins{h}as not yet \kgsins{been} established. Lidov-Kozai cycles with tidal friction \citep[KCTF;][]{Eggleton2001} \kgsins{have been} hypothesized but require a long dynamical timescale and only operate for certain mutual orbital inclinations. Thus, \kgsins{KCTF is} suspected to only be capable of producing a fraction of the total known population of close binaries. 

\kgsins{Recent} simulations of newborn triple systems \citep[e.g.,][]{Reipurth2012} \kgsins{have} found binary orbital hardening occurring early in the systems' evolution, whereby compact triple systems dynamically unfold into wider hierarchical structures. In this scenario, triple systems find themselves in the tight binary -- wide tertiary configuration well before evolving onto the main sequence. These findings have been corroborated by recent population synthesis work \citep[e.g.,][]{Moe2018a}, which found $\sim$60\% of close binaries form in this manner, with additional energy dissipation arising from primordial gas-disk interactions in the binary. 

\kgsdel{The ability to establish} \kgsins{Making progress in determining} the respective rates of these different mechanisms requires a greater number of \kgsins{binary-star} systems with well known periods, distances, and ages, \kgsins{and for which the presence of a tertiary companion is well established}. 
\kgsins{In addition, the fundamental result of an increased tertiary occurrence among the tightest binaries \citep{Tokovinin2006} should ideally be reproduced with independent samples, given its importance in motivating and constraining these questions.}

% The primary targets of these surveys involve spectroscopic binaries (SBs), in part due to their ubiquity and utility. Starting in the early 2000s, working towards a complete picture of binary formation, however, required an explanation for the increasing observational evidence of tightly bound SBs, with separations on the order of tenths of an AU. As theoretical work argued in situ formation to be increasingly unlikely, alternative theories included dynamical orbital evolution via kozai cycles with tidal friction (KCTF, \citet{Eggleton2001}) induced by gravitational interactions with a third star. 

% High spatial resolution campaigns in the near-IR have the ability to maximize detections of faint distant companions. In 2006, Tokovinin et al.\ observed 62 solar-type spectroscopic binaries (SBs) with the NACO adaptive optics system. They found virtually all ($\sim$96\%) short-period (P $<$ 3$^{d}$) systems had tertiary companions.\ Surprisingly, no further efforts to reproduce this canonical result have been reported in the literature thus far. 

In this paper, we seek to \kgsins{test the reproducibility of} the cornerstone result of \citet{Tokovinin2006} and explore the evidence of potential evolutionary pathways for triple systems. In Section~\ref{sec:data}, we describe our sample of observed binaries and list some of their fundamental physical properties. We also describe the source and nature of our data along with data reduction and processing procedures. In Section~\ref{sec:companion}, we describe our search for additional companions to our sample using published catalogs and with {\it Gaia} DR2, in some cases offering additional confirmation of our Robo-AO multiplicity identifications. In Section~\ref{sec:results}, we report the results of our Robo-AO SB multiplicity survey and compare with complementary Robo-AO observations of {\it Kepler} EBs. In Section~\ref{sec:discussion}, we discuss the \kgsins{implications} of our results on current star-formation and binary-evolution theory. Finally, in Section~\ref{sec:summary}, we conclude with a brief summary of our findings.

%% file: Section_Files/Binary_Sample.tex
\section{Data and Methods}\label{sec:data}

\subsection{Spectroscopic Binary Star Samples Used}

The goals of this study include examining multiplicity fractions as a function of inner binary period and comparing the derived properties of our identified tertiaries with known multiples from the literature. Therefore, our Robo-AO target list required a large sample of spectroscopic binaries (SBs) with known orbital periods and distances. To this end, we use the \citet{Troup2016} sample of main-sequence Apache Point Observatory Galactic Evolution Experiment (APOGEE) stars identified with stellar and substellar companions, which includes 178 systems within 1~kpc. In addition, we also include the \citet{Torres2010} sample of benchmark eclipsing binaries (EBs which are also SBs). These 94 systems have known orbital periods and distances \citep{StassunTorres:2016} as well as component masses and radii derived to an accuracy of 1--3\%. 

\begin{figure}[!htp]
    \includegraphics[width=\linewidth,trim=65 20 65 65 ,clip]{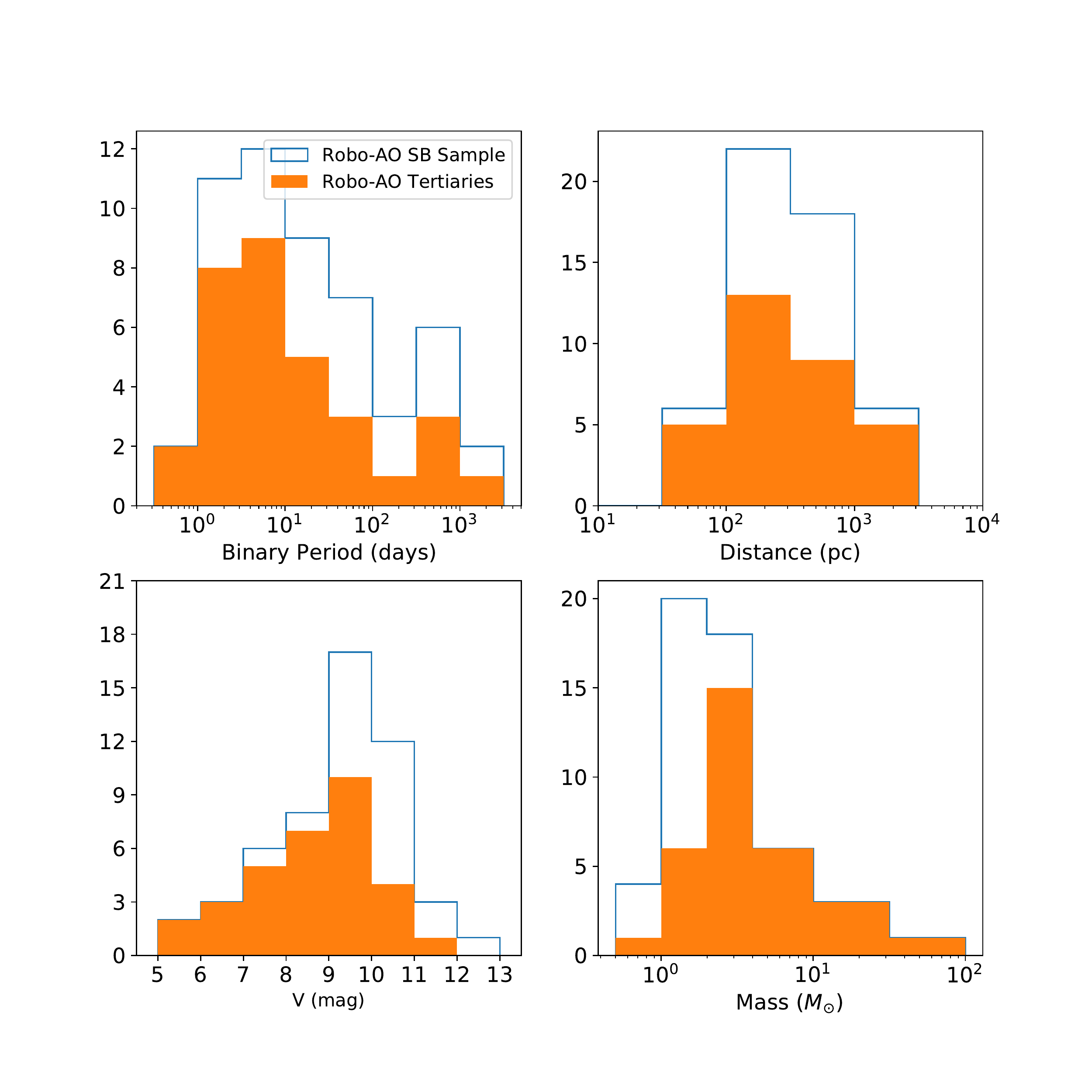}
    \caption{Representative histograms of stellar properties for the Robo-AO SB sample (unfilled) and the subset with identified tertiary companions (filled). We note the mass histograms refer to binary mass except for our \citet{Troup2016} sources, which only have known mass values for the primary. The bias against detection of very low-mass companions (lower-right panel) is a reflection of the sensitivity limit of our data for faint sources (lower-left panel). }
    \label{fig:starproperties}
\end{figure}

These two samples constitute the master list of SBs in our Robo-AO survey, of which we have observed 33 EBs from the \citet{Torres2010} sample and 43 targets from the \citet{Troup2016} sample, with orbital periods of the combined sample spanning the range 0.3--1880~d. Although all 43 Troup sources are radial velocity variable, we verify a clean sample of SBs by requiring the derived minimum mass of the companion to be greater than that of a brown dwarf (0.013\(\textup{M}_\odot\)). We also conservatively require the significance of the RV variations in sigma units to be greater than 10. This identifies 27 \citet{Troup2016} SBs (thus 33 + 27 = 60 SBs in total) for our sample. We do not consider the remaining 16 \citet{Troup2016} stars in our analysis in Sections \ref{sec:companion} or \ref{sec:results}, but for the benefit of future studies we report any companions that we identified in our Robo-AO imaging in Table \ref{tab:RobonotSBs}. All of the targets in our sample have parallaxes reported in the {\it Gaia\/} second data release (DR2) with distances in the range 40--2200~pc. Graphical summaries of some basic parameters for our Robo-AO SB sample are displayed in Figure~\ref{fig:starproperties}. Their properties and identifications are listed in Table~\ref{tab:Robofullsample}. 

\begin{table*}
\centering
\caption{Robo-AO SB Sample} \label{tab:Robofullsample} 
\begin{tabular}{cccccccc}
\hline \hline
Name & R.A. (deg) & Dec. (deg) & $P_{\rm bin}$ (d) & $V$ (mag) & Multiplicity Flag\tablenotemark{a} & Sample ID\tablenotemark{b} \\
\hline
V432 Aur & 84.38524 & 37.08689 & 3.082 & 8.05 & T & Stassun \& Torres - EB\\
WW Aur & 98.11313 & 32.45482 & 2.525 & 5.82 & T & Stassun \& Torres - EB \\
HS Aur & 102.82702 & 47.67335 & 9.815 & 10.05 & B & Stassun \& Torres - EB \\
HD 71636 & 127.48465 & 37.07095 & 5.013 & 7.90 & U & Stassun \& Torres - EB \\
KX Cnc & 130.69238 & 31.86242 & 31.22 & 7.19 & T & Stassun \& Torres - EB \\
2MASS J10352794+2512348 & 158.86626 & 25.20971 & 25.09 & 9.73 & B & Troup - SB \\
2MASS J11452973+0159347 & 176.3738 & 1.99299 & 1033.875 & 9.96 & B & Troup - SB \\
\hline
\end{tabular}
\footnotetext[1]{Multiplicity flag: B = binary, T = triple, U = undetermined (Section 2.2).}
\footnotetext[2]{Sample ID: Stassun \& Torres - EB = EBs from \citet{StassunTorres:2016}; Troup - SB = SBs from \citet{Troup2016}}
\tablecomments{The full table is available in the electronic version of the Journal. A portion is shown here for guidance regarding its form and content.}
\end{table*}

% \begin{table*}
% \centering
% \caption{Robo-AO SB Sample} \label{tab:Robofullsample} 
% \begin{tabular}{cccccccc}
% \hline \hline
% Name & R.A. (deg) & Dec. (deg) & $P_{\rm bin}$ (d) & $V$ (mag) & Strehl Ratio (\%) & Multiplicity Flag\tablenotemark{a} & Sample ID\tablenotemark{b} \\
% \hline
% V432 Aur & 84.38545 & 37.08674 & 3.082 & 8.05 & 1.68 & T & Stassun \& Torres \\
% WW Aur & 98.11327 & 32.45490 & 2.525 & 5.82 & 2.02 & T & Stassun \& Torres \\
% HS Aur & 102.82699 & 47.67338 & 9.815 & 10.05 & 1.58 & B & Stassun \& Torres \\
% 2MASS J07381910+2251256 & 114.57962 & 22.85714 & 23.906 & 12.499 & 6.85 & B & Troup - SB \\
% HD 71636 & 127.48463 & 37.07097 & 5.013 & 7.9 & 1.74 & U & Stassun \& Torres \\
% KX Cnc & 130.69255 & 31.86260 & 31.22 & 7.19 & 3.77 & Q & Stassun \& Torres \\
% 2MASS J10352794+2512348 & 158.86639 & 25.20971 & 25.09 & 9.733 & 1.27 & B & Troup - SB \\

% \hline
% \end{tabular}
% \footnotetext[1]{Multiplicity flag: B = binary, T = triple, Q = quadruple, U = undetermined (Section 2.2).}
% \footnotetext[2]{Sample ID: Torres = \citet{Torres2010}, Troup - SBs = SBs from \citet{Troup2016}}
% \tablecomments{The full table is available in the electronic version of the Journal. A portion is shown here for guidance regarding its form and content.}
% \end{table*}

For an independent test sample, 
we also made use of the published Robo-AO observations of {\it Kepler\/} EBs from \citet{Law2014}, 
originally drawn from the master sample of 
%To search for extrasolar planets, Robo-AO has also been systematically observing every 
{\it Kepler\/} objects of interest (KOIs). 
%a subset of which are EBs, thus providing an independent test sample. 
To enable a direct comparison with our Robo-AO sample, we trim the initial sample of 1065 KOI EBs as follows. To remove potential EB false positives, we required a minimum primary eclipse depth of 1~mmag, as fit by the {\it Kepler\/} EB pipeline's {\tt polyfit} algorithm \citep{Prsa2008}. Similarly, we also required all systems to have a successful ``morphology" classification (between 0 and 1) as output by the Local Linear Embedding (LLE) of the {\it Kepler\/} EB pipeline \citep{Matijevic2012}. We excluded all targets fainter than the faint limit of our sample ($V<12.5$) by 
%requiring $K_p < 13$. Lastly, we used a derived temperature-color relation (cite?) to roughly 
converting their {\it Kepler\/} magnitudes to $V$ via the published {\it Kepler\/} color-temperature relation. This also requires the EBs to have nominal effective temperatures listed in the {\it Kepler\/} Input Catalog. 
This leaves a remaining sample of 109 KOI EBs.

Of these 109, 52
%of these systems 
have Robo-AO observations, of which 22 
%systems having 
have clear determinations of the presence or absence of a companion 
%or lack there of 
(Table~\ref{tab:KOIEBsample}). These identifications correspond to multiple Robo-AO KOI survey efforts taken at the Palomar 1.5m telescope, including \citet{Law2014}, \citet{2016AJ....152...18B} \citet{Ziegler2017}, and \citet{2018AJ....155..161Z}. 

\begin{table*}
\centering
\caption{Robo-AO KOI EB Sample} \label{tab:KOIEBsample}
\begin{tabular}{cccccc}
\hline \hline
KOI & R.A. (deg) & Dec. (deg) & P$_{bin}$ (d) & V (mag) & Multiplicity Flag\tablenotemark{a} \\
\hline
5774 & 283.86634 & 47.22828 & 2.4275 & 10.834 & T \\
5993 & 285.14501 & 39.18703 & 4.2647 & 13.0 & T \\
971 & 286.01929 & 48.86677 & 0.5331 & 7.642 & B \\
6109 & 287.83336 & 39.22124 & 22.9135 & 11.911 & T \\
1728 & 288.97163 & 44.62465 & 12.7319 & 12.069 & B \\
2758 & 289.74253 & 39.26713 & 253.3623 & 12.168 & B \\
1661 & 290.73807 & 39.91969 & 1.8955 & 11.601 & T \\
\hline
\end{tabular}
\footnotetext[1]{Multiplicity flag: B = binary, T = triple.}
\tablecomments{The full table is available in the electronic version of the Journal. A portion is shown here for guidance regarding its form and content.}
\end{table*}

%In this paper, we work with \citet[][hereafter %\citetalias{Stassun2019}]{Stassun2019}

%% file: Section_Files/Imaging.tex
\subsection{Robo-AO Imaging}\label{sec:roboao}

%\subsection{Data Acquisition and Image Processing}

Robo-AO is an autonomous laser adaptive optics system stationed at the Kitt Peak 2.1m Telescope from November 2015 to June 2018. Robo-AO has a field size of 36\arcsec on a side with a pixel scale of 35.1 milli-arcsec per pixel \citep{Jensen-Clem2017}. High spatial resolution images of the target stars were taken between November 2017 and June 2018. We observed 76 unique targets in the $i'$ bandpass (6731 - 8726\AA) with 90-s exposures. Over the 90 second exposure, about 773 individual frames are generated. Target images were initially processed with the Robo-AO ``bright star'' pipeline as described in \citet{Law2014}.\ 

While this pipeline was appropriate for the majority of our Robo-AO SB images, we note five faint SB systems (2MASS IDs: J07381910, J16515260, J19301035, J19305116, and J19412976) that were not properly reduced. As noted in \citet{Jensen-Clem2017}, the images in these cases failed to correctly center the PSF, leading to a single noticeably bright pixel in the center of the image. We remove these targets from the sample to avoid biasing our multiplicity analysis in Section 4, leaving 55 SBs. To maximize our detection of tertiaries at small angular separations, the images were further processed by the ``high contrast imaging'' pipeline described in \citet{Jensen-Clem2017}. 

This pipeline first applies a high-pass filter on a 3\farcs5 frame windowed on the star of interest to lessen the contribution from the stellar halo. A synthetic point spread function (PSF) is then subtracted, generated via Karhunen-Lo'{e}ve Image Processing \citep[KLIP, a principal component analysis algorithm that whitens correlated speckle noise;][]{2012ApJ...755L..28S}. A representative sample of PSF diversity is achieved using a reference library of several thousand 3\farcs5 square high pass filtered frames that have been visually vetted to reject fields with more than one point source. This technique of Reference star Differential Imaging \citep[RDI;][]{2009ApJ...694L.148L} results in the final PSF-substracted image, from which we make our multiplicity determinations.

\begin{figure*}[!ht]
    \centering
    \includegraphics[width=\textwidth]{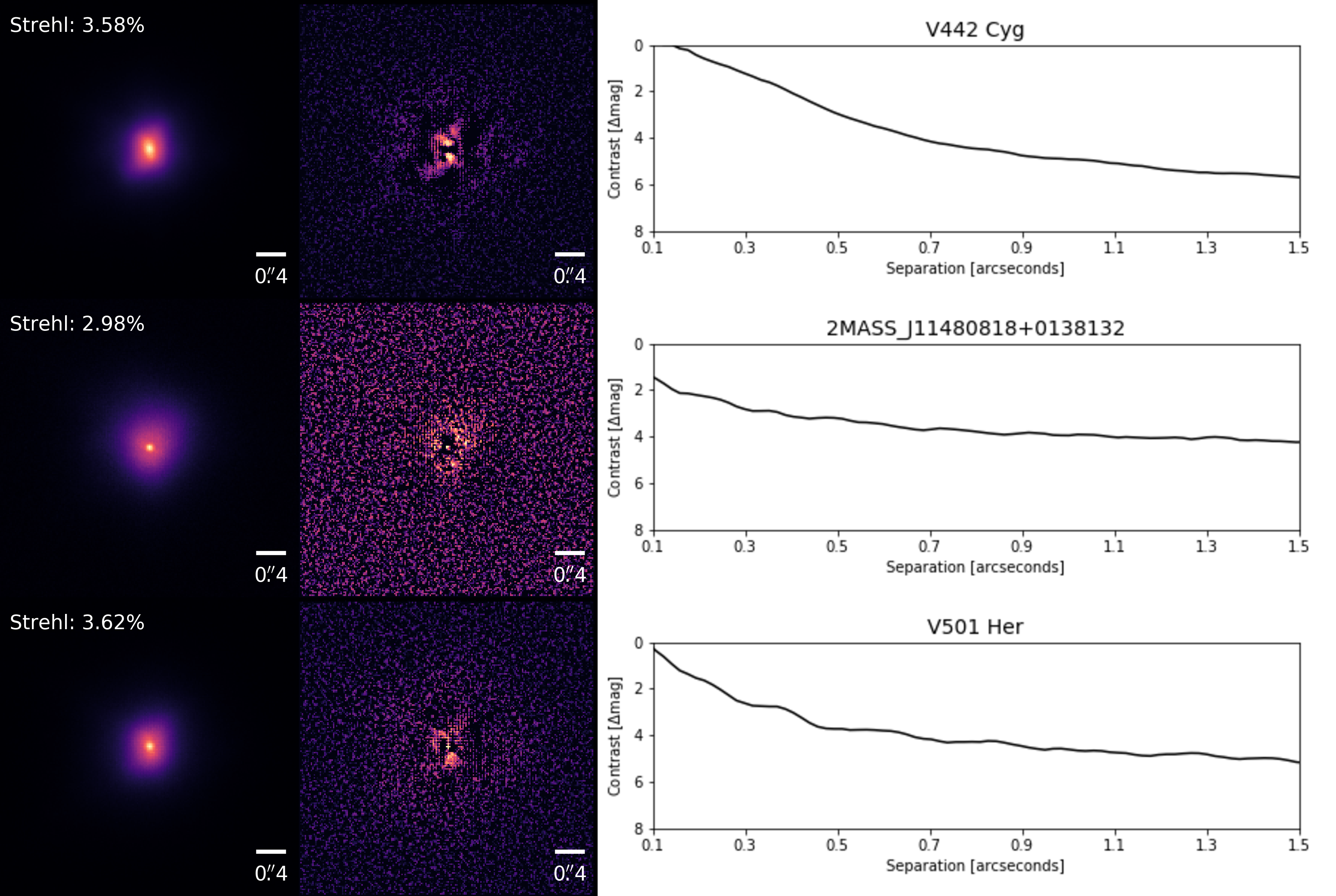}
    \caption{Top Panel: Robo-AO image \kgsins{(left)}, PSF subtracted image \kgsins{(middle)}, and constrast curve (right) of V442 Cyg; residuals \kgsins{resolve} a tertiary companion to the \kgsins{unresolved spectroscopic} binary. Middle Panel: Robo-AO Image, PSF subtracted image, and contrast curve of 2MASS J11480818+0138132; residuals indicate no detected tertiary companion. Bottom Panel: Robo-AO Image, PSF subtracted image, and contrast curve of V501 Her; residuals indicate unclear evidence of a companion.}
    \label{fig:AOimages}
\end{figure*}

Measurement of multiple star systems’ position angles and separations also require a precise astrometric solution for the optical instrument. Nightly observations of densely populated globular clusters are used to establish and update this solution as described in Section 6 of \citet{Jensen-Clem2017}. After each science target is fully reduced, the pipeline also produces a 5$\sigma$ contrast curve by simulating companions and properly correcting for algorithmic throughput losses using the Vortex Image Processing (VIP) package \citep{2017AJ....154....7G}. Examples of our Robo-AO images pre- and post-processing, along with their corresponding contrast curves, are shown in Figure~\ref{fig:AOimages}.

% The resulting PSF-subtracted images achieve minimal full-width at half-maximum of $\sim$0\farcs1; in practice, we achieve contrasts of up to $\sim$4 magnitudes at 0\farcs5 separations and up to $\sim$2.5 magnitudes at 0\farcs2 separations; for our median seeing of 1\farcs5 we, on average, detect companions at separations of 0\farcs36. 

The expected Robo-AO error budget and performance is summarized in Table 2 of \citet{Jensen-Clem2017}. At our observed Strehl ratios of a few percent, we expect a delivered FWHM of $\sim$0\farcs15. The majority of our SB sample have unambiguous detections or non-detections of a tertiary companion upon visual inspection of their PSF subtracted images. We assign these cases a multiplicity flag of ``T", or "B" for cases of undetected companions.. In a small number of cases, we observe residuals that weakly imply a quadruple but are suspected to be artifacts from the PSF-subtraction process. We assign these cases a multiplicity flag of ``T(Q?)" in Figures~\ref{fig:PSFpresubmosaic1}--\ref{fig:PSFpresubmosaic3}. Lastly, three cases have difficult to interpret residuals for which we are unable to determine identifications for (Figure~\ref{fig:AOimages}, lower right) and are thus assigned a multiplicity flag of ``U". Our analysis in Section~\ref{sec:results} excludes all unclear cases, resulting in a remaining sample of 52 SBs. 

% \begin{figure*}[!ht]
%     \centering
%     \includegraphics[width=0.825\textwidth]{Figures/merge_contrast.jpg}
%     \caption{Top Panel: Robo-AO \kgsins{(left)} and PSF subtracted \kgsins{(right)} image of 2MASS J08145689+3208572; residuals \kgsins{resolve} a tertiary companion to the \kgsins{unresolved spectroscopic} binary. Middle Panel: Robo-AO Image and PSF subtracted image of 2MASS J11480818+0138132; residuals indicate no detected tertiary companion. Bottom Panel: Robo-AO Image and PSF subtracted image of V501 Her; residuals indicate unclear evidence of a companion.}
%     \label{fig:contrastcurves}
% \end{figure*}

\begin{table*}
\centering
\caption{Robo-AO and Gaia Identified Tertiaries} \label{tab:Robotertiaries} 
\begin{tabular}{ccccccccc}
\hline \hline
Name & P$_{bin}$ (d) & Sep.\ (")\tablenotemark{a} & Dist.\ (pc) & Proj.\ Sep.\ (AU)\tablenotemark{a} & Mass (M$_\odot$)\tablenotemark{b} & $\sim$log (P$_{3}$) (yr)\tablenotemark{a} & GOF\_AL\tablenotemark{c} & D\tablenotemark{d} \\
\hline
V432 Aur & 3.082 & 0.12 & 127.1 & 15.6 & 2.3 & 1.2 & 5.45 & 0.0 \\
WW Aur\tablenotemark{f} & 2.525 & 0.13 & 90.9 & 12.2 & 3.8 & 0.9 & 29.27 & 32.69 \\
KX Cnc & 31.22 & 0.12 & 49.3 & 6.1 & 2.3 & 0.5 & 11.84 & 0.0 \\
2MASS J11480818+0138132\tablenotemark{e} & 400.82 & 119.1 & 235.0 & 27988.1 & 1.1 & 6.2 & 64.57 & 223.42 \\
2MASS J12260547+2644385 & 300.056 & 0.22 & 127.4 & 27.7 & 0.8 & 1.8 & 33.07 & 80.09 \\
BH Vir & 0.817 & 0.19 & 149.3 & 28.1 & 2.2 & 1.5 & 7.75 & 0.0 \\
AD Boo\tablenotemark{f} & 2.069 & 0.25 & 195.4 & 49.6 & 2.6 & 1.9 & 25.38 & 13.91 \\
2MASS J15044648+2224548 & 45.595 & 0.19 & 67.2 & 12.9 & 1.2 & 1.2 & 22.45 & 14.48 \\
2MASS J15212898+6722473 & 767.137 & 0.20 & 255.3 & 51.6 & 1.2 & 2.1 & 15.69 & 10.56 \\
RT CrB & 5.117 & 0.21 & 401.1 & 85.8 & 2.7 & 2.2 & 8.9 & 0.0 \\
V335 Ser\tablenotemark{f} & 3.45 & 0.22 & 202.3 & 43.9 & 4.1 & 1.7 & 17.46 & 3.97 \\
2MASS J16063131+2253008 & 1316.38 & 0.16 & 86.2 & 13.6 & 1.2 & 1.2 & 40.87 & 45.54 \\
2MASS J16100946+2312212\tablenotemark{f} & 75.887 & 0.26 & 437.1 & 114.5 & 1.2 & 2.6 & 20.31 & 3.71 \\
WZ Oph & 4.184 & 0.18 & 156.9 & 28.7 & 2.4 & 1.5 & 9.31 & 0.0 \\
V2365 Oph\tablenotemark{f} & 4.866 & 0.20 & 254.4 & 51.1 & 3.0 & 1.9 & 15.29 & 4.27 \\
V2368 Oph & 38.327 & 0.18 & 217.6 & 38.1 & 4.9 & 1.6 & 20.24 & 10.79 \\
LV Her & 18.436 & 0.25 & 374.2 & 94.3 & 2.4 & 2.3 & 11.3 & 0.0 \\
V624 Her\tablenotemark{f} & 3.895 & 0.18 & 140.6 & 25.9 & 4.2 & 1.4 & 28.95 & 21.52 \\
KIC 8410637 & 408.324 & 0.16 & 1266.7 & 200.1 & 2.8 & 2.8 & 13.55 & 0.0 \\
DI Her\tablenotemark{f} & 10.55 & 0.16 & 650.5 & 103.4 & 9.7 & 2.1 & 24.51 & 7.28 \\
FL Lyr & 2.178 & 0.20 & 135.0 & 26.5 & 2.2 & 1.5 & 10.11 & 0.0 \\
V885 Cyg\tablenotemark{f} & 1.695 & 0.21 & 964.0 & 206.3 & 4.2 & 2.7 & 21.82 & 5.05 \\
KIC 3858884 & 25.952 & 0.19 & 552.8 & 106.7 & 3.7 & 2.3 & 13.12 & 0.0 \\
V380 Cyg\tablenotemark{f} & 12.426 & 0.14 & 1061.9 & 150.8 & 18.4 & 2.2 & 66.94 & 103.34 \\
V453 Cyg & 3.89 & 0.17 & 1518.0 & 255.0 & 25.0 & 2.5 & 10.54 & 0.0 \\
V478 Cyg & 2.881 & 0.18 & 2222.8 & 406.8 & 30.4 & 2.7 & 8.21 & 0.0 \\
MY Cyg & 4.005 & 0.17 & 250.5 & 41.8 & 3.6 & 1.7 & 9.17 & 0.0 \\
V442 Cyg & 2.386 & 0.19 & 340.5 & 64.3 & 3.0 & 2.0 & 7.68 & 0.0 \\
Y Cyg\tablenotemark{f} & 2.996 & 0.16 & 1735.2 & 277.6 & 35.5 & 2.4 & 22.83 & 7.64 \\
CG Cyg & 0.631 & 0.22 & 97.5 & 21.5 & 1.8 & 1.4 & 8.94 & 0.0 \\
V364 Lac & 7.352 & 0.15 & 410.5 & 61.2 & 4.6 & 1.9 & 14.33 & 0.0 \\
BK Peg\tablenotemark{e} & 5.49 & 104.3 & 318.6 & 33257.6 & 2.7 & 6.1 & 11.56 & 0.0 \\
\hline
\footnotetext[1]{We note the reported quantities in these columns are upper limits.}.
\footnotetext[2]{For Stassun \& Torres SBs, masses are sourced from the detached eclipsing binary catalogue (DEBCat, \citet{2015ASPC..496..164S}).}
\footnotetext[3]{{\it Gaia\/} Astrometric Goodness of Fit in the Along-Scan direction.}
\footnotetext[4]{{\it Gaia\/} significance of the Astrometric Excess Noise.}
\footnotetext[5]{Wide tertiary companion identified by Gaia CPM analysis (Section 3.1).}
\footnotetext[6]{Multiple status astrometrically confirmed by Gaia (Section 3.2).}
\end{tabular}
\end{table*}

We do not resolve individual components at the raw image level for the majority of our Robo-AO observations (Figure~\ref{fig:PSFpresubmosaic1}--\ref{fig:PSFpresubmosaic3}). Many of the triples revealed in the PSF-subtracted images, however, show clear elongation compared to the point-source like observations of SBs with no detected companions. While the residuals from this process allow us identify the presence of a companion, it cannot be used to reliably measure flux contrasts. For each identified tertiary, we only measure an upper limit to its angular separation and a position angle by measuring the positions of the two peaks (i.e., the central unresolved SB and the tertiary companion) in the PSF subtracted images, carefully accounting for the slight difference in x and y pixel scales \citep[noted in the appendix of][]{Jensen-Clem2017}. These are then translated into physical upper limit projected separations using the {\it Gaia\/} DR2 distance to each system. We note the masses for each target have been sourced from the literature and correspond to the binary mass, except for our \citet{Troup2016} sources, which only have known mass values for the primary. These masses are used to derive a rough upper limit estimate of the tertiary period. We report its logarithm along with our measured angular and projected separations for all of our multiple systems in Table 3. We note that our period estimates, however, can significantly vary from the true period in cases where the projected separation is much different than the true semi-major axis of the system. 

% \sout{Finally, we also derive approximate magnitude contrasts, measured with simple aperture photometry on the two peaks (i.e., the central unresolved SB and the tertiary companion) in the PSF subtracted images. Our measured position angles and magnitude contrasts for the resolved companions in our sample are reported in Table 6 for incorporation into the Washington Double Star (WDS) catalog. Duplicate entries correspond to first the tertiary and then the quaternary companion of the system. Similarly, contrasts limits are reported at multiple separations for the Rob-oAO SBs that we do not resolve companions for in Table 7.}

%and their
%
%The 
%distribution 
%of our estimated tertiary periods 
%is shown in Figure~\ref{fig:perioddist}. 

The resolution and sensitivity limits of our Robo-AO observations hinder our ability to detect
%short ($P_{3} \lesssim 30$~yr) 
long period ($P_{3}\gtrsim 10^4$~yr)
tertiaries. The 3\farcs5 frame centered on each of our sources corresponds to a maximum detectable angular separation of 1\farcs75. For a typical tertiary in our sample, this corresponds to a tertiary period of $\sim$9200~yr. Thus, very long period ($P_{3}\gtrsim 10^4$~yr) tertiaries generally do not fall within our Robo-AO field of view. 

%Considering the typical mass \((2.2~\textup{M}_\odot\)) and distance (350 pc) for a binary in our sample, the Robo-AO minimum resolution of $\sim$0\farcs2 corresponds to an estimated tertiary period of $\sim$60~yr. 

%corresponds to an angular separation of 0.136". 
%As can be seen in Table~\ref{tab:Robotertiaries}, we observe angular separations of $\sim$0.3", rarely approaching 0.25". Thus, typical ($P_{3} \lesssim100$) tertiaries are not resolved in our imaging data. 

%% file: Section_Files/Companion.tex
\section{Catalog Search for Additional Companions}\label{sec:companion}

As the most sensitive probe of parallaxes and proper motions, {\it Gaia\/} provides a powerful opportunity to search more comprehensively for additional companions to our Robo-AO SB sample. Resolved common proper motion (CPM) companion matches have the ability to corroborate, or in some cases refine, our prior multiplicity determinations. Additionally, astrometric quality information of unresolved companions can similarly serve to confirm Robo-AO-identified multiples. As a final companion check, we also query the WDS catalog for prior multiplicity information on our Robo-AO targets.

\subsection{Gaia Common Proper Motion Candidates}

We begin by crossmatching our Robo-AO SBs with the {\it Gaia\/} DR2 catalog to obtain both parallax and proper motion information for each target. We then query {\it Gaia\/} DR2 to list all targets found within a 5$'$ aperture centered on each Robo-AO binary. The aperture size chosen searches for any wide companions outside of our 3\farcs5 Robo-AO field of view while minimizing those with separations that are unlikely to be bound. For both this larger sample as well as our Robo-AO-Gaia crossmatch, we apply the suggested corrections to the published {\it Gaia\/} DR2 magnitudes following \citet{2018A&A...616A...4E} and \citet{2018A&A...619A.180M}, depending on the magnitude range considered. For the brighter (\textit{G} $<$ 11.5) stars in both samples, we also apply a proper motion correction due to the inertial spin of the Gaia DR2 proper motion system \citep{2018IAUS..330...41L}.

Our wide search returns 49343 sources, 6810 of which do not have parallax or proper motion information. With the remaining 42533 sources, we follow a similar procedure to the Gaia companion candidate cuts chosen by \citet{2019AJ....157...78J} to find wide co-moving binaries. In our case, we choose more lenient fractional error cuts on parallax (20\%) and proper motion (50\%) to maximize our chances of identifying fainter companions while still ensuring reasonably high quality measurements. This vetting reduces the total number of candidate companions to 5332.

% To exclude the noisest systems, we also require non-negative parallaxes as well as fractional parallax errors of 20\% or less for each of our sources, reducing this sample to 2541 sources. We note our reported results in the following are not sensitive to this cutoff. 

To check for CPM companions that are likely to be physically associated, we then compare the parallax and proper motion information of these candidates with their corresponding Robo-AO counterpart. We require their respective parallax and proper motion measurements to agree within 2.5$\sigma$, with $\sigma$ being the larger error of the two. For the remaining 10 candidates, we estimate their projected separations, with distances derived from the error weighted arithmetic mean parallax of each pair. Again, we note these estimates can vary significantly from the true semi-major axis of the orbit. As a final requirement, we set a generous separation threshold of 50000 AU to exclude pairs that are most likely not physically bound while recognizing the approximate nature of our derivations.

The result is the identification of three CPM companion candidates. We report a faint (\textit{G} $\sim$ 18.9) companion to RT CrB at an angular separation of 9\farcs5. We change its multiplicity designations from triple to quadruple and make a note of its non-hierarchical (2 + 1 + 1) configuration. We also report a faint (\textit{G} $\sim$ 18.9) companion to BK Peg at an angular separation of 104\farcs4 and a (\textit{G} $\sim$ 11) companion to 2MASS J11480818+0138132 at an angular separation of 119\farcs1. We change their multiplicity designations from binary to triple.

Lastly, we return to the subsample of 7108 sources that {\it Gaia\/} does not record parallax or proper motion information for. Although these sources cannot be vetted for association, we check for those that have separations from their Robo-AO counterpart near Gaia's resolution limit of 1\arcsec. We find one match to 2MASS J12260547+2644385, a confirmation of the wide ($\sim$1\arcsec) triple companion confirmed in our Robo-AO imaging. Considering the other Robo-AO system with a relatively wider companion in the Robo-AO field of view, we find the ($\sim$1\arcsec) triple companion for CG Cyg is not detected by {\it Gaia}. However, considering the proximity to {\it Gaia\/}'s resolution limit as well as the rarity of chance alignments at these separations, we choose to leave its multiplicity designation unchanged. 
% \sout{We note the quaternary companion for KX Cnc is also not detected, expected from its close separation ($\sim$0\farcs23).}

% After changed search criteria, below no longer found
% We report one faint (\textit{G} $\sim$ 17.7) wide companion candidate to V478~Cyg at an angular separation of 170\farcs8, which would make V478~Cyg a quadruple. However, we keep its designation as a triple considering that a common proper-motion companion at such wide separation ($\sim$400,000~AU) is potentially unbound.

\subsection{Gaia Astrometric Candidates}

Significant deviations from {\it Gaia}'s astrometric fit, manifesting as large values of the Astrometric Goodness of Fit in the Along-Scan direction (GOF\_AL) and the significance of the Astrometric Excess Noise (D), hint at the presence of unresolved companions. In particular, \citet{2018RNAAS...2...20E} finds a cutoff of GOF{\_}AL $>$ 20 and D $>$ 5 to best separate confirmed binaries from confirmed singles. To confirm the source of the astrometric noise, we explore its dependence as a function of inner binary separation using the slightly relaxed criterion of GOF{\_}AL $>$ 15 and D $>$ 3.

As we show in Figure 3, Gaia does not register significant astrometric noise (threshold marked by the dotted line) for the majority of our Robo-AO SBs, which typically have inner binary separations $\lesssim$ 1.5~mas (left of dashed line). For those that do register astrometric noise in this regime, we find all (10) systems also have a companion identified in our Robo-AO imaging. 
%We note in Table~\ref{tab:Robotertiaries} these RoboAO identified companions that also show this astrometric signal. We note the other three systems in this regime with lower, non-zero recorded astrometric noise (below dotted line) do not pass our astrometric criterion and do not have companions identified from our RoboAO imaging.
We conclude that systems whose central SBs are tighter than $\sim$1.5 mas are not noticed as multiples by {\it Gaia}, unless they also possess a wider tertiary.

In contrast, we find the vast majority (12/14) of our wider SBs ($\gtrsim$1.5~mas, right of dashed line) register significant astrometric noise values (above dotted line), even if they do not possess an identified tertiary companion.
%of 0.15~mas or greater; In this regime, however, further evidence of a companion through our astrometric criterion is not as clear, with our RoboAO imaging not finding companions for 6 SBs that pass (2MASS IDs: J11452973, J12261982, J12590520, J16100901, J16294740, and J17130501) . 
In this regime, {\it Gaia\/} is likely detecting the photocenter motion of the relatively wide SBs, regardless of whether there is a wider tertiary. 
%It is unclear, however, why Gaia does not record astrometric noise for three of these wider SBs (right of dashed, below dotted).

% \textbf{We find the vast majority (10/12) of our wider SBs ($\gtrsim$2~mas) register astrometric noise values of 0.15~mas or greater; presumably, {\it Gaia} is detecting the photocenter motion of these relatively wide SBs.}

%2MASS J16100901{+}2443358, and 2MASS J16294740{+}3941054

% \textbf{We note 6 SBs (2MASS IDs: J11452973, J12261982, J12590520, J16100901, J16294740, and J17130501) originally identified in our RoboAO imaging as having no companions %also 
% just barely pass the GOF{\_}AL and D criteria.
% %This subset of four do not appear to be stronger outliers in this metric. 
% Considering the lack of confirmation in our RoboAO data along with their unconvincing astrometric status, we do not change their original identifications.} 

A similar analysis for our Robo-AO KOI EB sample also confirms 
%6 multiple sources and identifies one binary as potentially having an unresolved companion.
our tertiary designations. The sole exception is
%For this binary (
KOI 6016, for which {\it Gaia} identifies significantly high values 
%for our criterion, with 
of GOF{\_}AL $=$ 166 and D $=$ 2095. Assuming a total binary mass of 2~M$_\odot$, we estimate the angular separation of the EB (0.677 mas) 
%and compare to 
to be much smaller than its measured astrometric excess noise (2.38 mas), 
%Given these values, we find the detected astrometric noise cannot correspond to the binary and 
thus we argue its source is an unresolved tertiary companion. We change its designation from binary to tertiary.

\begin{figure}[!htp]
    \centering
    \includegraphics[width=\linewidth]{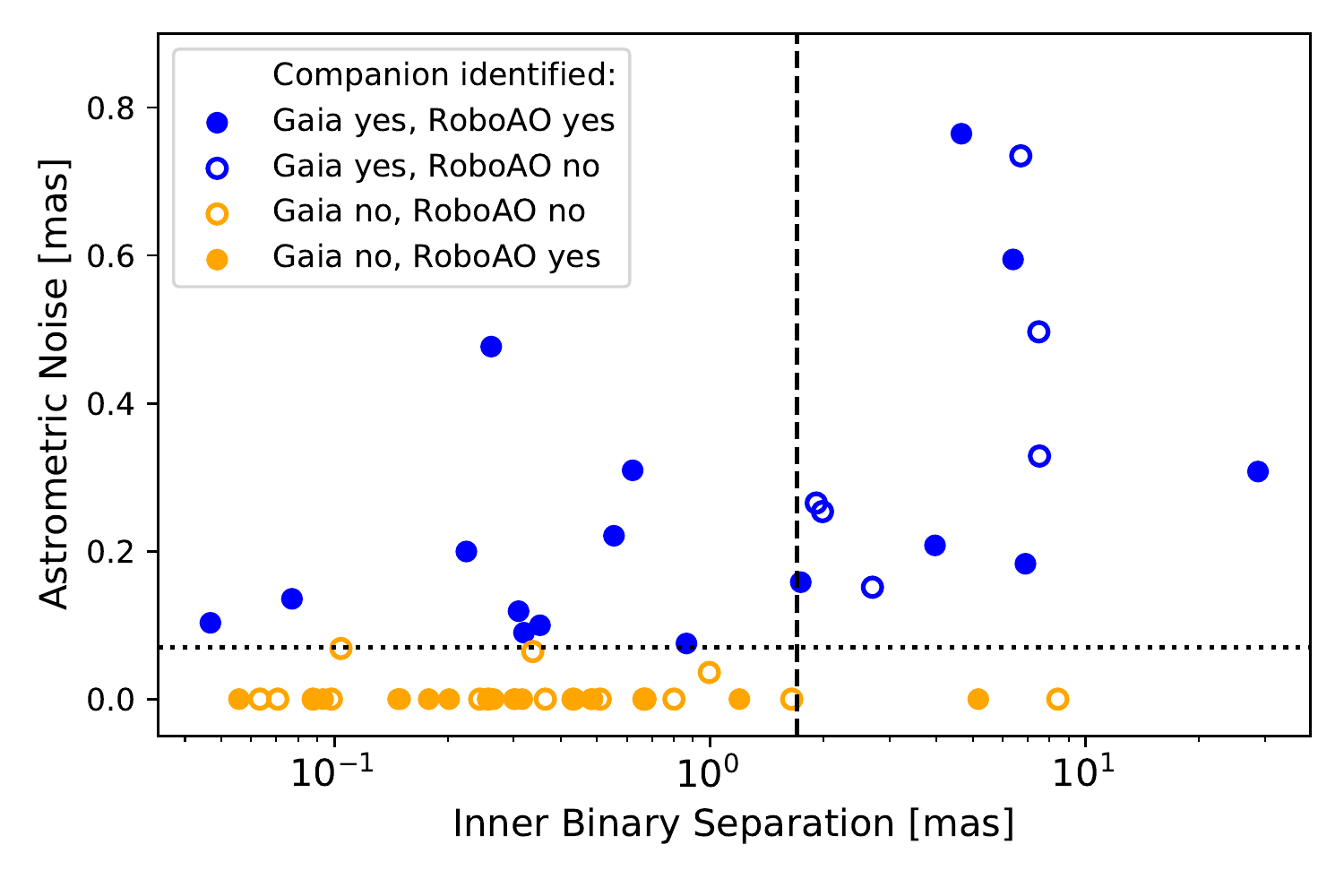}
    \caption{Gaia recorded astrometric noise for our Robo-AO SBs as a function of inner binary separation. Systems that (do not) pass our Gaia astromeric criterion (GOF{\_}AL $>$ 15 and D $>$ 3) are shown in blue (orange). Filled (unfilled) dots correspond to cases where our Robo-AO imaging (did not) identified a companion within 3.5\arcsec. The dotted line denotes the minimum astromtric noise observed for systems which pass our astrometric criterion. The dashed line separates our tighter (left) and wider (right) Robo-AO SBs.}
    \label{fig:gaiaastrometric}
\end{figure}

\subsection{Washington Double Star Catalog Comparison}

In this section, we carefully compare our observations to those listed in the WDS catalog. We find 8 of our Robo-AO SB targets have entries and describe them individually below. 

\paragraph{2MASS J11480818+0138132}
The most recent companion identification listed in WDS entry 11482+0136 refers to observations in 2015 by the Garraf Astronomical Observatory (OAG). Its CPM wide pairs (WP) survey identified\footnote[1]{The OAG CPMWP catalog can be found at \url{https://www.oagarraf.net/Comunicacions/OAG\%20CPM/GWP\%20CATALOG\%20EQUATORIAL\%20ZONE\%202016_ASCII__1.0.txt}.} a \textit{V}$\sim$11.1 companion at a separation of 119\farcs.1 with a position angle of 334\degree. While this target is outside of the field of view of our Robo-AO imaging, our {\it Gaia} CPM analysis (see Section 3.1) also identifies this wide companion.

\paragraph{2MASS J12260547+2644385}
WDS entry 12261+2645 refers to a singular speckle-interferometric observation that identified a \textit{V}$\sim$9.7 companion at a separation of 1\farcs1 with a position angle (PA) of 167\degree (ASCC number 684901 in \citet{2015AJ....150...16G}). Our Robo-AO observations find 2MASS J12260547+2644385 to be a triple, with a tertiary companion at an upper limit separation of 1\farcs11 and a PA of 346\degree. Given our convention of measuring PA with respect to the SB (346\degree-180\degree=166\degree), we find these results to be in agreement and confirm this companion.

\paragraph{2MASS J16063131+2253008}
The most recent companion identification listed in WDS entry 16065+2253 refers to Pan-STARRS observations from 2015 as detailed by \citet{2014ApJ...792..119D}. They report a companion separation of 35\farcs5 at a PA of 22\degree. The proposed source, 2MASS J16063229+2253337, is resolved by \textit{Gaia} but faint (\textit{G}$\sim$19.3). {\it Gaia} reports a parallax, R.A.\ proper motion, and Dec.\ proper motion of 11.734 $\pm$ 0.418 mas, $-$84.088$\pm$ 0.538 mas/yr, and 90.03$\pm$0.571 mas/yr. In comparison, {\it Gaia} reports 11.601$\pm$0.075 mas, $-$88.263$\pm$0.105 mas/yr and 72.873$\pm$0.124 mas/yr for 2MASS J16063131+2253008. Given in particular the difference in Dec.\ proper motion, our {\it Gaia} CPM analysis does not find this pair to be associated.

\paragraph{2MASS J16074884+2305299}
The most recent companion identification listed in WDS entry 16078+2306 refers to a 2004 observation \citep{2015ApJS..219...12A} that identified a \textit{V}$\sim$13.3 companion at a separation of 12\farcs2 with a position angle (PA) of 81\degree. A source at this separation does not fall into our Robo-AO field of view. {\it Gaia} resolves one other source within 15\arcsec, Gaia DR2 1206535963616734976, and reports a parallax, R.A.\ proper motion, and Dec.\ proper motion of 0.735$\pm$0.02 mas, 3.58$\pm$0.03 mas/yr, and $-$9.18$\pm$0.03 mas/yr. In comparison, {\it Gaia} reports 2.04$\pm$0.036 mas, $-$3.48$\pm$0.04 mas/yr and $-$9.38$\pm$0.05 mas/yr for 2MASS J16063131+2253008. These sources are clearly not physically associated.

\paragraph{V2368 Oph}
WDS entry 17162+0211 refers to a singular 1985 speckle-interferometric observation from the Center for High Angular Resolution Astronomy (CHARA) \citep{1987AJ.....93..688M}. They reported a separation of 0\farcs136 at a PA of 69\degree. As detailed in the auxiliary WDS notes, the WDS entry was recalled after repeated attempts at confirmation by McAlister but later restored. While the most recent effort to confirm this companion \citep{2018MNRAS.473.4497R} was unsuccessful, our Robo-AO imaging detects a companion at an upper limit separation of 0\farcs175 at a PA of 68\degree. Given the close agreement in separation and PA, we tentatively confirm this companion but encourage continued monitoring of this clearly complex case.

\paragraph{V624 Her}
The most recent companion identification listed in WDS entry 17443+1425 refers to a 2015 Gaia DR1 observation \citep{2018JDSO...14..503K} that identified a \textit{V}$\sim$11.75 companion at a separation of 40\arcsec with a position angle (PA) of 151\degree. {\it Gaia} DR 2 reports a parallax, R.A.\ proper motion, and Dec.\ proper motion of 7.1119 $\pm$ 0.0628 mas, $-$2.271$\pm$ 0.098 mas/yr, and 15.134$\pm$0.096 mas/yr for V624 Her. In comparison, {\it Gaia} DR2 reports 0.7206$\pm$0.0457 mas, $-$0.446$\pm$0.074 mas/yr and $-$3.962$\pm$0.070 mas/yr for WDS 17443+1425B. These sources are clearly not physically associated.

\paragraph{V478 Cyg}
The most recent companion identification listed in WDS entry 20196+3820 refers to a 2015 listing in the Webb Deep-Sky Society's Double Star Section Circulars (DSSC)\footnote[2]{The DSSC catalog can be found at \url{https://www.webbdeepsky.com/dssc/dssc23.pdf}}. The listing, a 2006 observation from UKIDSS DR6 identifies a \textit{V}$\sim$14.5 companion at a separation of 3\farcs61 with a position angle (PA) of 258\degree. A source at this separation does not fall into our Robo-AO field of view. {\it Gaia} resolves one other source within 5\arcsec of V478 Cyg but shows the pair to have discrepant parallaxes. We opt to retain our original identification.

\paragraph{CG Cyg}
The most recent companion identification listed in WDS entry 20582+3511 refers to two speckle-interferometric observations in 2014 that identified a \textit{V}$\sim$12 companion at a separation of 1\farcs1 with a position angle (PA) of 310\degree \citep{2015AJ....150..151H}. Our Robo-AO observations find CG Cyg to be a triple, with a tertiary companion at an upper limit separation of 1\farcs16 and a PA of 313\degree. Given the close agreement in separation and RA, we confirm this companion.\\

% Out of this subsample, 4 entries refer to wide ($>$1\arcsec) sources that are seen neither in our Robo-AO images nor by {\it Gaia}, likely because they are faint (e.g., three of them have \textit{K} $\sim$ 18). We do not change our RoboAO designations for these sources.

% \textbf{Three entries of this subsample are clearly ruled out by the parallax and proper motion measurements of {\it Gaia}. As an example of one of these three, the entry for V624 Her refers to a visual companion (WDS 17443+1425B) separated by 40\arcsec.\ {\it Gaia} reports a parallax, R.A.\ proper motion, and Dec.\ proper motion of 7.1119 $\pm$ 0.0628 mas, $-$2.271$\pm$ 0.098 mas/yr, and 15.134$\pm$0.096 mas/yr for V624 Her. In comparison, {\it Gaia} reports 0.7206$\pm$0.0457 mas, $-$0.446$\pm$0.074 mas/yr and $-$3.962$\pm$0.070 mas/yr for WDS 17443+1425B. These sources are clearly not physically associated. 2 entries refer to companions with parallax and proper motion measurements that are more similar but only one (the companion candidate to 2MASS J11480818+0138132) passes our Gaia CPM criterion (Section 3.1). We note these cases in Tables 5 and 6 and suggest updates to the WDS catalog given the information in {\it Gaia} DR2.} 

%The remaining 3 entries list nearer ($\lesssim$1\farcs3) companions, all of which are confirmed by our Robo-AO imaging.

Given the faintness of the two wide companions we identified in Section 3.1 (\textit{G}$\sim$19), it is not surprising the WDS catalog does not list entries for RT Crb or BK Peg. Similarly, the remaining 43 Robo-AO SBs do not have entries in the WDS catalog, likely a consequence of the difficulty of finding visual companions at the proximity of the (upper limit) separations of Robo-AO identified companions ($\leq$0\farcs25).

%% file: Section_Files/Results.tex
\section{Results}\label{sec:results}

To summarize our analysis in Sections \ref{sec:data}--\ref{sec:companion}, 
%resulting in our final determinations. 
out of our initial sample of 55 SBs, we obtain Robo-AO determinations in 52 cases. After our catalog search for additional CPM and astrometric candidate companions, we report a final tally of 20 binaries, 31 triples, and 1 quadruple system.

%\subsection{Tertiary Incidence versus Binary Orbital Period}

\begin{figure}[!ht]
    \includegraphics[width=\linewidth,trim=10 0 30 25,clip]{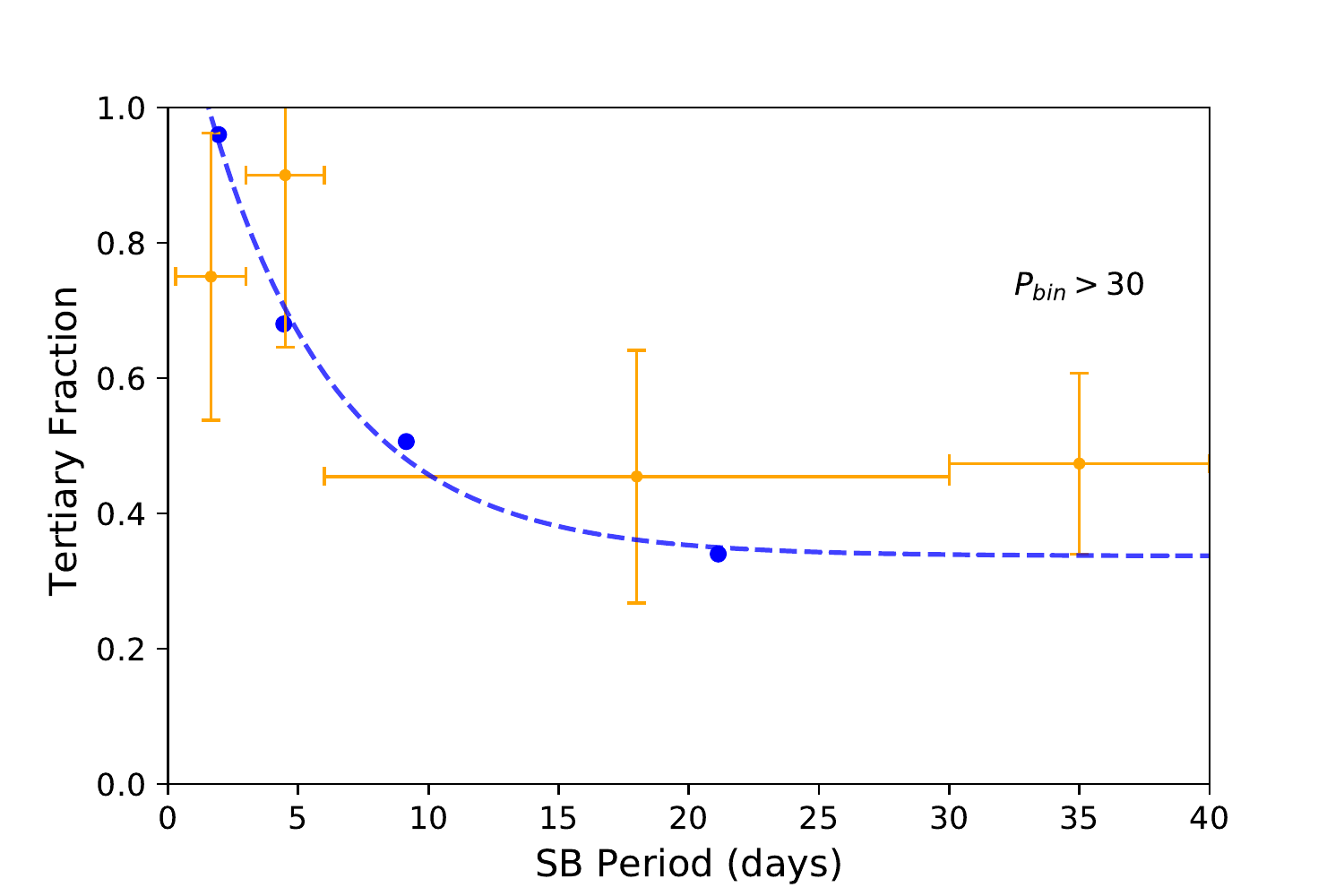}
    \caption{Fraction of SB systems with a tertiary companion as a function of binary period (orange). \kgsins{The final bin represents all SBs with $P_{\rm bin} > 30^d$.} %In red, we show our revised tertiary fractions after applying incompleteness corrections.
    In blue, we overplot the tertiary fractions found in \citet{Tokovinin2006}, fit with a decaying exponential. Individual error bars correspond to the adjusted Poisson error for a multinomial distribution.} %We list binomial probabilities above each period bin, assuming our overall tertiary fraction ($\sim$43\%) as our tertiary companion probability. We also list the binomial probabilities for our sample when binned more coarsely.
    \label{fig:tertiaryfracvsbperiod}
\end{figure}

%\subsubsection{New Robo-AO Results}
We begin by exploring the dependence of tertiary companion frequency on inner SB period for our Robo-AO SBs. Our 52 systems are sorted into SB period bins, with bin edges of 0, 3, 6, and 30 days (Figure~\ref{fig:tertiaryfracvsbperiod}, orange). Systems with periods greater than 30~d are grouped together (rightmost point). We find a trend of increasing incidence of tertiary companions toward shorter period SBs (orange points), with 90\% of ($3^{d} < P_{\rm bin} < 6^{d}$) SBs having a tertiary companion compared to $\sim$47\% of the longest period ($P_{\rm bin} > 30^{d}$) SBs. In the shortest period bin ($P_{\rm bin} < 3^d$), we find a slightly decreased fraction of 75\% relative to the next bin ($3^{d} < P_{\rm bin} < 6^{d}$). We report the resultant tertiary fraction we derive for each period bin with its error in Table~\ref{tab:terfractable}. 

\begin{table}%{llcc}
\centering
\caption{Derived Tertiary Fractions\label{tab:terfractable}} 
\begin{tabular}{|c|cc|}
\hline 
%\tablehead{
%\colhead{Sample 1} & \colhead{Sample 2} & \colhead{KS Statistic} & %\colhead{P-value}
%}
Period Bin & N & \textit{f} \\ 
\hline
%\startdata
%\cutinhead{Tertiary Period Distributions}

\multicolumn{3}{c}{Robo-AO SBs} \\
\hline

$P_{1} < 3$ & 12 & 0.75 $\pm$ 0.21 \\

$3 < P_{1} < 6$ & 10 & 0.9 $\pm$ 0.25 \\

$6 < P_{1} < 30$ & 11 & 0.45 $\pm$ 0.19 \\

$P_{1} > 30$ & 19 & 0.47 $\pm$ 0.14 \\
%\cutinhead{Tertiary vs.\ Binary Period Distributions}

\hline
\multicolumn{3}{c}{Robo-AO KOI EBs} \\
\hline

$P_{1} < 3$ & 5 & 0.96 $\pm$ 0.4 \\

$3 < P_{1} < 6$ & 3 & 0.68 $\pm$ 0.47 \\

$6 < P_{1} < 30$ & 8 & 0.51 $\pm$ 0.18 \\

$P_{1} > 30$ & 6 & 0.34 $\pm$ 0.24 \\
%($P_{3} < 10^6$ yr) & ($P_{1} > 10$ yr) \\
%\cutinhead{Tertiary-to-Binary Period Ratio Distributions}
\hline
\end{tabular}
% \footnotetext[1]{This study; clipped at detection limits in all calculations.}
% \footnotetext[2]{\citet{Tokovinin2018} (MSC)}
% \footnotetext[3]{\citet{Tokovinin2006}.}
% \footnotetext[4]{\citet{Raghavan2010}.}
\end{table}

%To account for incompleteness, we apply corrections to the tertiary frequency in each SB period bin, as follows. Because our Robo-AO observations are not sensitive to tertiary companions within $\sim$0\farcs25 of the SBs, we assume that our main observational bias is against short period tertiaries. For the typical distance of our target sample (see Section~\ref{sec:data}), this corresponds to $P_{3} \lesssim 30$~yr. 
%We use the 
%subset of these systems found in the 
%\citet{Tokovinin2006} sample as defining the relative fraction of tertiaries with $P_{3} > 30$~yr 
%We assume their detection limits to be lower limits 
%and divide our raw fractions by that fraction. 
%of identified multiple systems in each binary period range.
%This increases the total inferred tertiary occurrence rate (Figure~\ref{fig:tertiaryfracvsbperiod}, red points), up to \kgsins{96\%} for our shortest period SBs ($P_{\rm bin} < 3^{d}$).

\kgsins{To compare with the canonical trend originally reported by \citet{Tokovinin2006}, in Figure~\ref{fig:tertiaryfracvsbperiod} we represent the \citet{Tokovinin2006} tertiary fractions with a simple exponential (note that this is not intended to represent a physical model). The tertiary fractions that we observe as a function of inner binary period are broadly consistent with the \citet{Tokovinin2006} result. We find this trend also extends to binary periods of $P_{\rm bin} > 30^d$, which are even longer than those considered by \citet{Tokovinin2006}.}

%\subsubsection{KOI EBs Observed by Robo-AO}

% Of the 
% %remaining 
% 108 EBs in the KOI sample (see Section~\ref{sec:data}), 48 
% %of these systems 
% have Robo-AO observations, of which 22 
% %systems having 
% have clear determinations of the presence or absence of a companion 
% %or lack there of 
% (Table~\ref{tab:KOIEBsample}). These identifications correspond to multiple Robo-AO KOI survey efforts taken at the Palomar 1.5m telescope, including \citet{Law2014}, \citet{2016AJ....152...18B} \citet{Ziegler2017}, and \citet{2018AJ....155..161Z}. Similar to our methodology above for identifying tertiary companions, the authors analyzed the PSF-subtracted images to look for stars within 4" of the central target. Our {\it Gaia} analysis (Section 3.1) rules out the wider ($\gtrsim$2-3\arcsec) identified candidate companions. Candidate companions within $\sim$1\arcsec of the targets are considered to be likely physically associated. The resulting tertiary fraction as a function of SB period 
% %for the Robo-AO KOI EB sample 
% is shown in Figure~\ref{fig:tertiaryfracKOI}. 

\begin{figure}[!ht]
    \includegraphics[width=\linewidth,trim=10 0 30 25,clip]{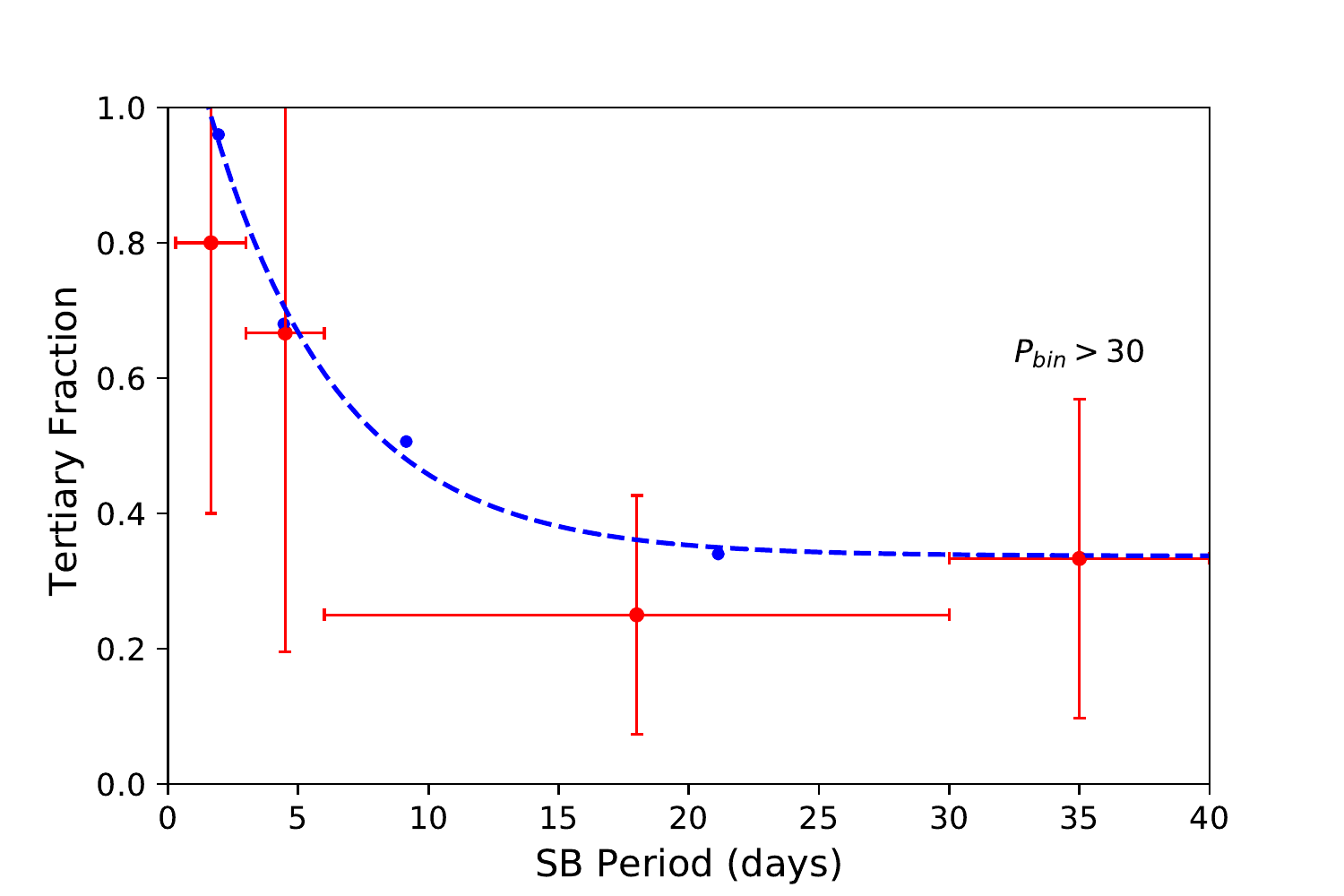}
    \caption{Same as Figure~\ref{fig:tertiaryfracvsbperiod}, but for KOI EBs observed by Robo-AO (red). The final bin represents all SBs with $P_{\rm bin} > 30^d$.}
    \label{fig:tertiaryfracKOI}
\end{figure}

Next, as outlined in Section 2.1, we use the KOI EBs observed by Robo-AO as an additional, independent test sample of this trend. Similar to our methodology above for identifying tertiary companions, these systems had their PSF-subtracted images analyzed to look for stars within 4" of the central target (references in Section 2.1). Our {\it Gaia} analysis (Section 3.1) rules out the wider ($\gtrsim$2-3\arcsec) identified candidate companions. Candidate companions within $\sim$1\arcsec of the targets are considered to be likely physically associated. The resulting tertiary fractions as a function of SB period 
%for the Robo-AO KOI EB sample 
are shown in Figure~\ref{fig:tertiaryfracKOI} and again reported with errors in Table~\ref{tab:terfractable}. 

Although there are only 10 KOI tertiaries among the SBs with $P_{\rm bin} < 30^{d}$, 
%to compare with, 
we find general agreement between our Robo-AO results (Figure~\ref{fig:tertiaryfracvsbperiod}) and the KOI EB sample Figure~\ref{fig:tertiaryfracKOI}. We again find a systematically higher fraction of tertiary companions with shorter period SBs, with both trends clearly consistent with the findings of \citet{Tokovinin2006}. 
%While we are unable to apply a suitable incompleteness correction, 
We again note this trend also seems to extend to binary periods of $P_{\rm bin} > 30^{d}$.

%% file: Section_Files/Discussion_summary.tex
\section{Discussion}\label{sec:discussion}
%Possible points for discussion {\bf [make sure that the evolutionary story, from compact configurations to widening tertiaries, comes across by the end of this discussion]}: 

% A principal aim of this paper was to test the reproducibility of the canonical result from \citet{Tokovinin2006}, namely that tertiary frequency is a strong function of inner binary orbital period. 
% Our new sample of Robo-AO observations of field SBs with known orbital periods
% is smaller (roughly half the size) than the \citet{Tokovinin2006} sample, limiting the statistical power of the comparison. 
% %our sample is potentially less representative of the true distribution of tertiary periods.
% Nonetheless, we do find broad consistency with the trend reported by \citet{Tokovinin2006}
% \kgsins{and we moreover find that the trend of decreased tertiary fraction at longer SB periods extends to even longer SB periods, $P_{\rm bin} > 30^d$ (Figure~\ref{fig:tertiaryfracvsbperiod}). This consistency with the \citet{Tokovinin2006} result is corroborated as well as with another independent sample of KOI EBs (Figure~\ref{fig:tertiaryfracKOI}).} 
\subsection{Tertiary-to-Binary Orbital Period Ratios}

\begin{figure}[!ht]
    \includegraphics[width=\linewidth,trim=5 0 25 30,clip]{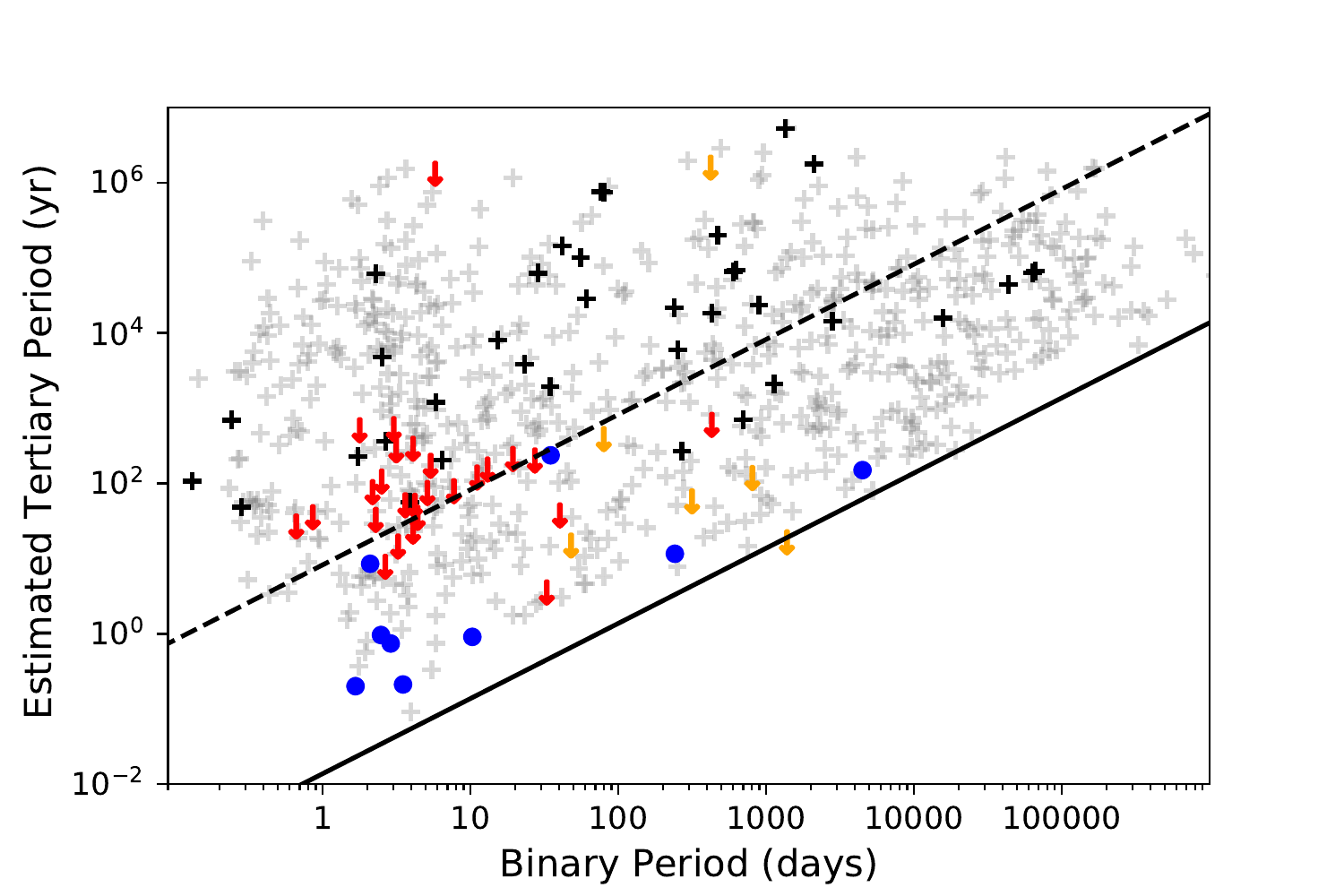}
    \caption{Estimated tertiary periods as a function of inner binary period.  Identified Robo-AO triples are marked as red \mbox{\citep{Torres2010}} and orange \mbox{\citep{Troup2016}} arrows, given our estimates are upper limits. For comparison, we plot the updated MSC triples \mbox{\citep{Tokovinin2018}} as grey crosses while overplotting the triples from the volume-limited \mbox{\citet{Raghavan2010}} sample as black crosses. For reference, the dashed line represents $P_3 = 10^{3.5} P_{1}$. The dynamical stability limit for triples ($P_{3} = 4.7P_{1}$) is shown by the full line. Known PMS ($\lesssim$30~Myr) triples are shown in blue.}
    \label{fig:stabilitycheck}
\end{figure}

The increasing number of known multiples with well-defined orbital periods has lead to a thorough probing of the distribution of tertiary orbital period as a function of inner binary period. In Figure~\ref{fig:stabilitycheck}, the large sample of known triples from the updated Multiple Star Catalog (MSC) \citep{Tokovinin2018} (grey crosses) and the smaller sample of triples from the volume-limited \mbox{\citet{Raghavan2010}} sample (black crosses) clearly indicate the wide range of permitted tertiary-binary period ratios. However, as noted in both \mbox{\citet{2014AJ....147...86T}} and \mbox{\citet{Tokovinin2018}}, the rarity of systems with small period ratios ($<$ 10), and in particular short tertiary periods ($P_{3} <$ 10~yr), is evident from the lack of points along the dynamical stability limit $P_{3} = 4.7P_{1}$ \mbox{\citep[solid line;][]{Mardling2001}} in the lower left corner.

In an effort to explore a potential correlation with age, we refer to the benchmark PMS EB sample \mbox{\citep{Stassun2014}}, of which 7 of the 13 are identified with or have evidence of a tertiary companion. Three of these systems \citep[RS Cha, TY CrA, and MML 53;][]{2013MNRAS.432..327W, Corporon1996, GomezMaqueoChew2019} have tertiaries with orbital solutions; their ages range from 3--15~Myr. A literature search for additional PMS tertiaries also reveals orbital solutions for V1200~Cen \mbox{\citep{Coronado2015}}, GW Ori \mbox{\citep{2017ApJ...851..132C}}, TWA 3A \mbox{\citep{2017ApJ...844..168K}}, V807 Tau \mbox{\citep{2012ApJ...756..120S}}, as well as TIC 167692429 and TIC 220397947 \mbox{\citep{2020MNRAS.tmp..465B}}.

The tendency for these PMS systems to lie at low tertiary-binary period ratios is evident in Figure~\ref{fig:stabilitycheck}. To probe the potential significance of this apparent difference relative to the older field population, we perform a two-dimensional, two-sided KS test between the PMS sample and the volume-limited \mbox{\citet{Raghavan2010}} sample, reporting a p-value $< 10^{-4}$. The difference is statistically significant, which corroborates the visual impression that indeed the PMS sample occupies a different distribution of tertiary-to-binary orbital periods than the field-age sample.

For our Robo-AO sample, we note the upper limits of our estimated tertiary periods (red and yellow arrows in Figure~\ref{fig:stabilitycheck}) 
%primarily cluster around 10$^{3.5}$ years, 
are consistent with the larger field-age samples of \citet{Tokovinin2018} and \citet{Raghavan2010}. 

% For our newly identified tertiaries, 15 of our observed \citet{Torres2010} EBs have estimated ages. The majority (12) are field binaries, with ages between 300 Myr -- 10 Gyr. Looking at their period ratios as a function of age does not reveal any obvious trends. We note, however, this small sample size does not allow for definitive conclusions.

% Among this older subset, we do not note a 
% %into three age groups and explore their tertiary-to-binary period ratios as a function of age. 
% trend of increasing period ratio as a function of age. However, with a sample size of only 15 systems, 
% %with the older systems appearing to have wider tertiary companions relative to their inner binaries.
% In other words, the systems appear to become less compact and more hierarchical as they evolve. 

%A Kendall's $\tau$ test on these 10 systems, however, reports this trend is not formally significant, with a p-value of 0.3.

%The majority (11) are significantly older with ages in the range of 300 Myr -- 10 Gyr than the minority (1--4 Myr). 
%We find no correlation, however, between tertiary-binary period ratio and age for the younger minority in this subset.
%with both groups exhibiting a wide range of period ratios. We note, however, not only the small number of pre-main-sequence systems but also the significant expected error of these ages (25-50\% or in some cases greater).

\kgsdel{However, with the inclusion of our shortest period binaries (P$_{\rm bin} < 3^{d}$), we find evidence of a flatter, and potentially decreasing within the error bars, distribution of tertiary fractions at this end. This result implies the non-negligible occurrence of mechanisms that create close binaries without the influence of a companion, such as significant interactions with a primordial circumbinary disk \citep{1991ApJ...370L..35A}. Alternatively, close dynamical interactions near stellar birth could potentially account for a fraction these cases, with a low mass companion being completely removed from the system. Subsequently, these companions, at wide separations, would also not be recovered in common proper motions searches.} 

\subsection{Hierarchical Unfolding}

Although tertiary-induced inner binary period shortening is evident, there is growing evidence that a single mechanism, such as Lidov-Kozai cycles with tidal friction (KCTF), cannot recreate the entire population of known close binaries \citep[e.g.,][]{Kounkel2019,Bate2019}. As has already been shown for individual cases \citep[e.g.,][]{GomezMaqueoChew2019}, we find that many of the well-characterized EBs in young ($<$30~Myr) triple systems have already achieved close separations ($\lesssim$0.1~AU). Evolution by the KCTF mechanism, which has an effective timescale on the order of $\sim$100~Myr, is unable to account for these ($<$30~Myr) systems. 

The alternative mechanism of hierarchical unfolding \citep[e.g.,][]{Reipurth2012}, in which the orbit of a newly bound wide (100--1000 AU) binary is shrunk by ejecting a third body into a distant orbit, is unable to achieve the close separation of a spectroscopic binary alone. However, because these interactions often result in a highly eccentric orbit for the binary, the additional dissipative interactions with the primordial gas expected in the disks of these young binaries also aid in the observed orbital decay. Recent population synthesis work \citep{Moe2018a} finds that $\sim$60\% of close binaries ($P_{\rm bin} < 10^{d}$) form during the PMS phase and in a compact configuration with a tertiary companion as a consequence of this process.

Known PMS triples with estimated orbits are notably compact and weakly-hierarchical, with significantly smaller tertiary-to-binary period ratios than the majority of known triples in the field (see Figure~\ref{fig:stabilitycheck}). The majority of our Robo-AO triples with known ages (12/15) are field-age ($\sim$1~Gyr) systems, and have thus had sufficient time to dynamically evolve. In comparison to the PMS systems, they all lie in the cluster of points with upper limit period ratios of $10^{3.5}$ or greater. Although it is not feasible to reconstruct the dynamical pathway taken by each system, as both hierarchical unfolding and KCTF are able to account for the current orbital configurations, their hierarchical nature is evident, with a range of hardened inner binaries ($P_{\rm bin} < 30^d$) and large tertiary-to-binary period ratios ($\geq 10^{3.5}$).

This apparent difference between the field-age and PMS sample could be equally explained from the standpoint of the tertiary companion or the inner binary. The simulations of \citet{Reipurth2012} find the most extreme wide systems take on the order of $\sim$100~Myr to fully unfold and thus it is possible some of the tertiaries in these PMS systems still have significant dynamical evolution to undergo. Alternatively, if the reservoir of primordial circumstellar disk gas is not yet exhausted, it is possible some of our PMS binaries are still in the process of hardening if young enough. Thus, the conditions for evolution via hierarchical unfolding with dissipative gas interactions appear to be in place for the youngest systems.

%Looking at their period ratios as a function of age does not reveal any obvious trends, but we note a sample of this size does not allow for definitive conclusions.

\begin{figure}[!ht]
    \includegraphics[width=\linewidth,trim=0 0 0 0,clip]{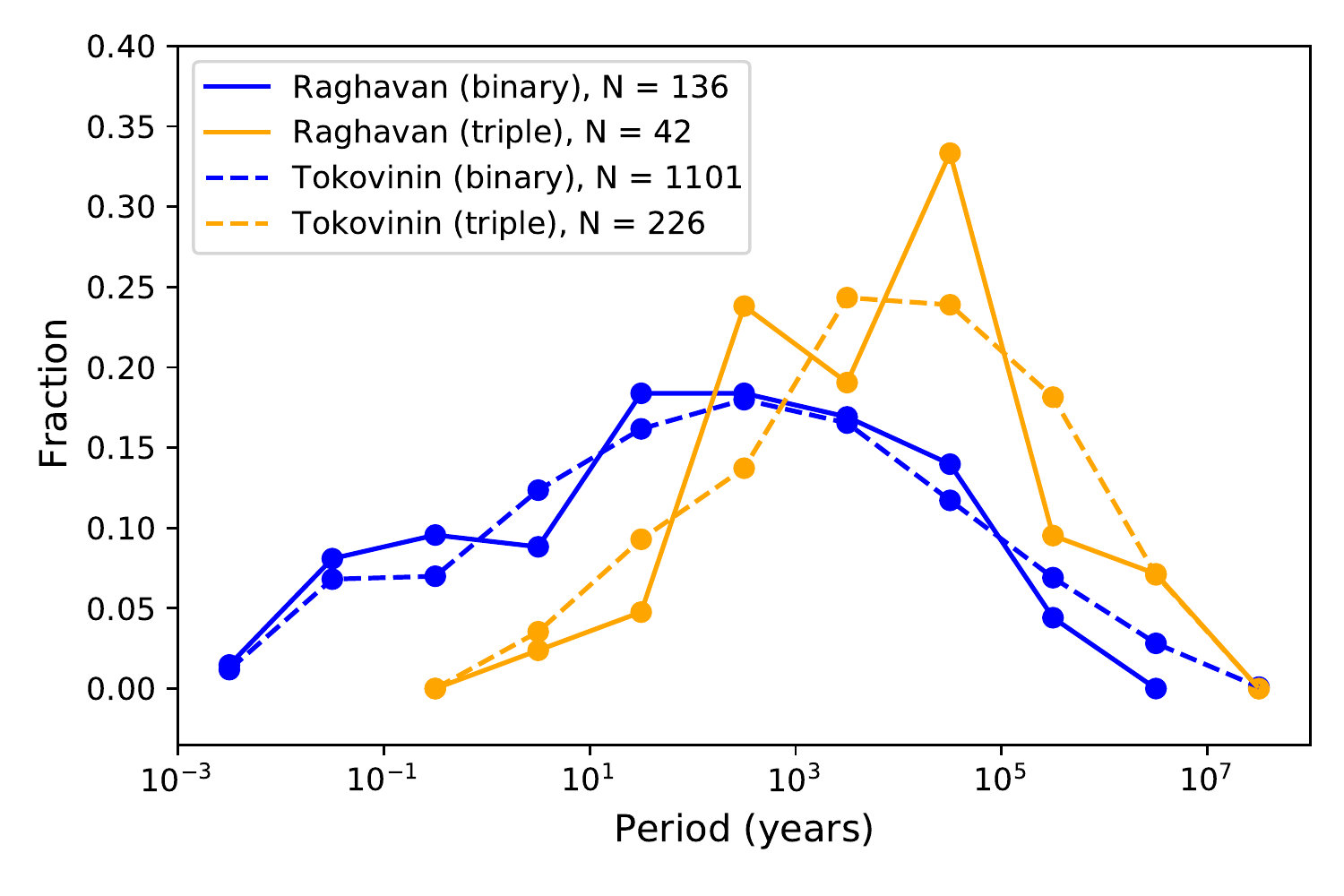}
    \caption{Distribution of orbital periods between the lone binaries (blue) and tertiaries (orange) from the volume-limited \citet{Raghavan2010} (solid) and \citet{2014AJ....147...86T} (dashed) samples.}
    \label{fig:ragperioddist}
\end{figure}

Another question is whether the 
%presence of a companion alters the underlying distribution of orbital periods, 
dynamical evolution of triple systems leads to tertiaries that are wider than would be expected to arise from the formation process alone. 
\citet{2014AJ....147...87T} argued that overall distributions of inner and outer orbital periods in multiple-star systems is consistent with dynamical sculpting that produces inner binaries populating the short-period part of the overall distribution and outer companions populating the long-period part of the overall distribution. 
But is there evidence that tertiaries come to reach wider separations than the widest binaries? 
%compared to that of lone binaries. 
To this end, we compare the period distributions of simple binaries versus triples within the volume-limited \citet{Raghavan2010} and \citet{2014AJ....147...86T} samples (Figure~\ref{fig:ragperioddist}). 

The upper bound of the appropriate period range for the comparison considers the longest period for which both binaries and tertiaries are observed ($\sim10^7$ years). Because inner binaries of triple systems cannot be as wide as their tertiaries, we do not extend the lower bound to the period for which the tightest tertiaries are observed (10 years); instead, we consider the longest period binaries in the triples identified in the larger volume-limited sample of \citet{2014AJ....147...86T}. We choose a generous lower bound of $10^3$ years, with only a small number of binaries (identified through CPM) exceeding this period.
   
Within this period range ($10^3 < P < 10^{7}$ years), we perform a two-sample Anderson-Darling test on the lone Tokovinin binaries versus tertiaries (dotted distributions in Fig.~\ref{fig:ragperioddist}) to probe the differences in the tails of these distributions. We do not find strong evidence (p-value $=$ 0.06) that tertiary companions find themselves at wider separations than their simple binary counterparts. We find a similar result when considering the Raghavan sample (solid distributions in Fig.~\ref{fig:ragperioddist}). There is therefore not strong evidence to suggest that the mechanisms governing the underlying populations of the widest binaries and tertiaries are distinct, consistent with the findings of \citet{2014AJ....147...87T}.

\section{Summary}\label{sec:summary}

In this paper, we have conducted robotic adaptive optics (Robo-AO) imaging for a sample of 60 spectroscopic binaries (SBs) with known periods and distances. For 52 individual sources, we identify companions, or lack thereof, through a visual inspection of their PSF subtracted images and search for additional candidates using {\it Gaia\/} parallaxes, proper motion, and astrometric information. Overall, we identify 31 tertiary systems and 1 quadruple systems.

% We examine the tertiary companion frequency as a function of inner SB period for not only our sample but also for the sample of {\it Kepler\/} eclipsing binaries (EBs) with published Robo-AO observations

A principal aim of this paper was to test the reproducibility of the canonical result from \citet{Tokovinin2006}, namely that tertiary frequency is a strong function of inner binary orbital period. For our Robo-AO identified tertiaries, we find higher fractions of tertiary companions around shorter period binaries, with 75\% of $P_{\rm bin} < 3^{d}$ systems having a tertiary companion compared to $\sim$47\% of the longest period ($P_{\rm bin} > 30^{d}$) systems, \kgsins{consistent with the findings of \citet{Tokovinin2006} and extending that result to even longer-period inner binaries}. A separate test sample of the smaller number of Robo-AO observed KOI EBs also shows this trend. The two samples, thus, appear to independently reproduce the canonical result of \citet{Tokovinin2006}. 

% Our new sample of Robo-AO observations of field SBs with known orbital periods is smaller (roughly half the size) than the \citet{Tokovinin2006} sample, limiting the statistical power of the comparison. Nonetheless, we find broad consistency with the trend reported by \citet{Tokovinin2006} \kgsins{and we moreover find that the trend of decreased tertiary fraction at longer SB periods extends to even longer SB periods, $P_{\rm bin} > 30^d$ (Figure~\ref{fig:tertiaryfracvsbperiod}). This consistency with the \citet{Tokovinin2006} result is corroborated as well as with another independent sample of KOI EBs (Figure~\ref{fig:tertiaryfracKOI}).} 

% A comparison of the distribution of tertiary periods to those of wide binaries suggests that the tertiaries extend to significantly wider separations than their simple binary counterparts, implying that the widest triple systems originate via mechanism(s) distinct from those that form the {\it in situ} distribution of binary separations. 

We have roughly estimated the tertiary period for each of our triples, exploring the dependence of their distributions on inner SB period and age. Although we are unable to determine the prominent mechanism at work for their dynamical evolution, we note all of our field-age Robo-AO triples find themselves in hierarchical configurations, with large ($P_{3}/P_{1} > 10^{3.5}$) tertiary-binary period ratio upper limits. In comparison, we find known young PMS triples are much more compact in comparison, with the conditions for hierarchical unfolding \mbox{\citep[e.g.,][]{Reipurth2012}} already in place. We find these results to be consistent with the recent population synthesis predictions of \mbox{\citet{Moe2018a}}.

%Our comparison of the distribution of tertiary periods to those of wide binaries in the volume-limited \citet{Raghavan2010} and \citet{2014AJ....147...86T} sample \textbf{does not find strong evidence} that tertiaries extend to significantly wider separations than their simple binary counterparts. \textbf{Given this, it is not clear that the} widest triple systems originate via mechanism(s) distinct from those that form the {\it in situ} distribution of binary separations.

% We do note, however, that the youngest tertiary systems find themselves in compact configurations with the inner binaries becoming hardened early on in their evolution. We find all of our field-age RoboAO triples have much greater tertiary-to-binary period ratios, consistent with the dynamical evolution expected by the recent population synthesis work of \citet{Moe2018a}.

% We find that several young PMS EB tertiaries seem to align well with the dynamical evolution pathways expected by the recent population synthesis work of \citet{Moe2018a}. We see that the youngest tertiary systems find themselves in compact configurations, with the inner binaries becoming hardened early on in their evolution before unfolding to a more hierarchical state, with greater tertiary-to-binary period ratios.  

Taken together, the results of this investigation can be interpreted through a framework in which stellar triples evolve from relatively compact configurations to increasingly hierarchical configurations, in which the hardest binaries arise almost exclusively through tertiary interactions, and in which the widest tertiaries arise through interactions with their inner binaries.

%% file: Section_Files/Acknowledgements.tex
\acknowledgments

This research has made use of NASA's Astrophysics Data System. This research has made use of the VizieR catalogue access tool and the SIMBAD database, operated at CDS, Strasbourg, France. This work made use of the IPython package \citep{PER-GRA:2007}, NumPy \citep{van2011numpy}, and SciPy \citep{2019arXiv190710121V}. This research made use of Astropy, a community-developed core Python package for Astronomy \citep{2013A&A...558A..33A} as well as matplotlib, a Python library for publication quality graphics \citep{Hunter:2007}. This research made use of ds9, a tool for data visualization supported by the Chandra X-ray Science Center (CXC) and the High Energy Astrophysics Science Archive Center (HEASARC) with support from the JWST Mission office at the Space Telescope Science Institute for 3D visualization \citep{2003ASPC..295..489J}. These acknowledgements were compiled using the Astronomy Acknowledgement Generator. Finally, the authors are most grateful to the anonymous referee, whose thorough and careful review substantially improved the manuscript.

This work has also made use of data from the European Space Agency (ESA) mission {\it Gaia} (\url{https://www.cosmos.esa.int/gaia}),processed by the {\it Gaia} Data Processing and Analysis Consortium (DPAC, \url{https://www.cosmos.esa.int/web/gaia/dpac/consortium}). Funding for the DPAC has been provided by national institutions, in particular the institutions participating in the {\it Gaia} Multilateral Agreement. This research made use of the cross-match service provided by CDS, Strasbourg. This research has made use of the Washington Double Star Catalog maintained at the U.S. Naval Observatory. The Robo-AO instrument was developed with support from the National Science Foundation under grants
AST-0906060, AST-0960343, and AST-1207891, IUCAA, the Mt. Cuba Astronomical Foundation, and by a gift from Samuel Oschin. We thank the NSF and NOAO for making the Kitt Peak 2.1-m telescope available as well as the observatory staff at Kitt Peak (KP) for their efforts to assist Robo-AO KP operations. The authors are honored to be permitted to use astronomical data observed on Iolkam Du’ag (Kitt Peak), a mountain with particular significance to the Tohono O’odham Nation. Robo-AO KP is a partnership between the California Institute of Technology, the University of Hawai‘i, the University of North Carolina at Chapel Hill, the Inter-University Centre for Astronomy and Astrophysics (IUCAA) at Pune, India, and the National Central University, Taiwan. Robo-AO KP was also supported by a grant from Sudha Murty, Narayan Murthy, and Rohan Murty, a grant from the John Templeton Foundation, and by the Mt. Cuba Astronomical Foundation. In particular, we thank the efforts of Christoph Baranec, Nicholas Law, Dmitry Duev, Reed Riddle, Maissa Salama, and Rebecca Jensen-Clem of the Robo-AO Kitt Peak team.


%% copy editing.  Individual instruments can be provided in parentheses,
%% after the keyword, but they will not be verified.

Facilities: \facility{KPNO: 2.1m (Robo-AO)}.